\documentclass[showpacs,showkeys, groupedaddress,twocolumn,
               prb, aps,floatfix,10pt]{revtex4-2}

\usepackage{natbib}

\usepackage[english]{babel}
\usepackage{fontenc}
\usepackage{graphicx}
\usepackage{amsmath}
\usepackage{epstopdf}
\usepackage{amssymb}
\usepackage{amsbsy}
\usepackage{amscd}
\usepackage{float}
\usepackage{xcolor}
\usepackage{tikz}
\usepackage{braket}
\usepackage{standalone}
\usepackage{bm}
\usepackage{siunitx}
\usepackage[normalem]{ulem} 
\usepackage{verbatim}
\usepackage{url}
\usepackage{silence}
\usetikzlibrary{positioning,calc}
\tikzset{>=latex}
\usepackage[verbose,hypertexnames=false,bookmarksopenlevel=1,filecolor=blue,
            linkcolor=blue,citecolor=blue,urlcolor=blue,pdfstartview=FitH,bookmarksopen,bookmarksnumbered,
            colorlinks,plainpages=false,linktocpage]{hyperref}
\usepackage{cleveref}
\newcommand\mb{\mathbf}

\begin{document}
\title{From Thermalization to Multifractality: Spin-Spin Correlation in Disordered $SU(2)$-Invariant 1D Heisenberg Spin Chains}
\author{Debasmita Giri}
\email[To whom correspondence should be addressed.  \\ Email: ]{debasmita.giri@physik.uni-regensburg.de}
\affiliation{Institute for Theoretical Physics, University of Regensburg, Regensburg, Germany}
\author{Julian Siegl}
\affiliation{Institute for Theoretical Physics, University of Regensburg, Regensburg, Germany}
\author{John Schliemann}
\affiliation{Institute for Theoretical Physics, University of Regensburg, Regensburg, Germany}

\begin{abstract}
We investigate spin correlations in one-dimensional $SU(2)$-invariant Heisenberg chains with exchange disorder for spin lengths $S=1/2$ and $S=1$.  In the weak-disorder regime, the eigenmodes of the spin-spin correlation matrix are delocalized, consistent with ergodic behavior. Under strong disorder, the system enters a quasi-localized multifractal phase characterized by exponentially decaying, dimer-like spin correlations. Finite-size scaling of the inverse participation ratios of the correlation-matrix eigenmodes yields a correlation dimension, $D_2\approx 0.37-0.39$, confirming the stability of a multifractal regime that is distinct from both the ergodic limit ($D_2=1$) and the fully localized limit ($D_2=0$). 
\end{abstract}
\maketitle

\section{Introduction}
Disorder and interactions in one-dimensional quantum spin chains give rise to rich non-ergodic phenomena that lie beyond the conventional eigenstate thermalization hypothesis (ETH)~\cite{deutschQuantumStatisticalMechanics1991,srednickiChaosQuantumThermalization1994,srednickiApproachThermalEquilibrium1999,nandkishoreManyBodyLocalizationThermalization2015,dalessioQuantumChaosEigenstate2016}.
In clean interacting systems, ETH predicts that expectation values for highly excited eigenstates behave as thermal ensembles so that local observables relax to their equilibrium values and two-point correlations decay rapidly with distance~\cite{nandkishoreManyBodyLocalizationThermalization2015,gogolinEquilibrationThermalisationEmergence2016}.
Anderson first demonstrated that in one dimension, even infinitesimal disorder is sufficient to localize noninteracting particles and halt diffusion~\cite{andersonAbsenceDiffusionCertain1958,mottAndersonTransition1975,adkinsThresholdConductionInversion1978}
In the decades since, the phenomenon of Anderson localization has become a cornerstone of condensed matter physics, with exact results in 1D and scaling theories in higher dimensions~\cite{abrahamsScalingTheoryLocalization1979,eversAndersonTransitions2008}.
The fate of localization becomes more subtle when taking interactions into account.
In the presence of sufficiently strong quenched disorder, many-body localization (MBL) can emerge: transport is frozen, entanglement growth is logarithmically slow, and local operators retain memory of their initial conditions even at infinite temperature~\cite{baskoMetalinsulatorTransitionWeakly2006,sierantManybodyLocalizationAge2025,imbrieManyBodyLocalizationQuantum2016,oganesyanLocalizationInteractingFermions2007,palManybodyLocalizationPhase2010,znidaricManybodyLocalizationHeisenberg2008,oganesyanEnergyTransportDisordered2009,berkelbachConductivityDisorderedQuantum2010,obrienExplicitConstructionLocal2016,serbynCriterionManyBodyLocalizationDelocalization2015,bardarsonUnboundedGrowthEntanglement2012,serbynLocalConservationLaws2013,vasseurQuantumCriticalityHot2015}.

Although there is still debate over the critical disorder strength of the transition~\cite{doggenManybodyLocalizationLarge2021,devakulEarlyBreakdownAreaLaw2015,maceMultifractalScalingsManyBody2019,sierantPolynomiallyFilteredExact2020,oganesyanLocalizationInteractingFermions2007,luitzManybodyLocalizationEdge2015,doggenManybodyLocalizationDelocalization2018,eversInternalClockManybody2023,selsThermalizationDiluteImpurities2023,colmenarezStatisticsCorrelationFunctions2019,pietracaprinaShiftinvertDiagonalizationLarge2018}, many studies suggest that once the disorder becomes strong enough, these systems will enter a fully localized phase~\cite{imbrieManyBodyLocalizationQuantum2016}.
On the contrary, studies on models with non‐Abelian symmetries have demonstrated that continuous symmetries, such as $SU(2)$, can obstruct the construction of local integrals of motion and thus hinder full localization~\cite{protopopovEffectSU2Symmetry2017,prelovsekAbsenceFullManybody2016,potterSymmetryConstraintsManybody2016,protopopovEffectSU2Symmetry2017,thomsonDisorderinducedSpinchargeSeparation2023,majidyNonAbelianSymmetryCan2023,protopopovNonAbelianSymmetriesDisorder2020,kozarzewskiSpinSubdiffusionDisordered2018,murthyNonAbelianEigenstateThermalization2023}.
In the random‐exchange Heisenberg chain, nearest‐neighbor couplings $J_i$ are drawn from a disorder distribution but the Hamiltonian retains full $SU(2)$ invariance.
This model emerged as a key testbed for this interplay.
In ref.~\cite{sieglImperfectManyBodyLocalization2023}, the authors have shown, via level statistics and gap‐ratio fluctuations, that neither GOE nor Poisson behavior occurs at strong disorder.
Rather, the model enters an “incompletely localized” regime.
Furthermore, in ref.~\cite{gaoSpectralEntanglementProperties2025} Gao and R\"omer have also shown, through entanglement diagnostics, that localization signatures are markedly weaker in the exchange-disordered model than in its random‐field counterpart.

While these spectral and entanglement diagnostics provide valuable global perspectives, they do not resolve the distribution of local two-spin operators or how two-point correlators are distributed across eigenstates in the strong disorder regime.
To bridge this gap, we present a comprehensive study of two‐point spin correlations of mid-spectrum eigenstates in both spin-$1/2$ ($N\leq 16$) and spin-$1$ ($N\leq 10$) random-exchange chains.
We focus on the $S^z_{\rm tot}=0$ sector for integer total spin and $S^z_{\rm tot}=1/2$ sector for half-integer total spin.
We begin by mapping out the full distribution of mid-spectrum (20$\%$ in the middle of the full spectrum) two-point correlators $\langle \hat{\bm{S}}_i\cdot\hat{\bm{S}}_{i+r}\rangle$, showing that at vanishing disorder these histograms are sharply peaked around zero for every separation $r$, consistent with infinite-temperature ETH, then develop small tails at moderate disorder, and discrete singlet$/$triplet (or multiplet) peaks at strong disorder.
Second, we extract the average absolute correlations $\langle |\hat{\bm{S}}_i\cdot\hat{\bm{S}}_{i+r}|\rangle$, which exhibits a flat profile for weak disorder and exponential decay in strong enough disorder.
Thirdly, we diagonalize the $N \times N$ spin-correlation matrix in each eigenstate, then obtain its eigenmodes, study their component distributions, and compute inverse participation ratios (IPRs).
We find that weak disorder yields extended modes with $\rm{IPR}\sim 1/N$, whereas strong disorder produces a plateau well above $1/N$ but below the ideal localized limit of $1$.
Finally, we perform finite‐size scaling of the disorder‐averaged IPR, $\langle \rm{IPR} \rangle \sim N^{-D_2}$, to extract the correlation‐dimensional exponent $D_2$.
In the ETH regime, $D_2\approx 1$~\cite{palManybodyLocalizationPhase2010}, confirming fully ergodic behavior.
In the strong‐disorder phase, $D_2\sim (0.37-0.39)$, well separated from both the ergodic and fully localized $(D_2=0)$limits, revealing a stable quasi-localized multifractal phase. These findings parallel previous studies of multifractality in Fock space in ref.~\cite{maceMultifractalScalingsManyBody2019}, where the multifractal dimensions $D_q$ were extracted from participation entropies computed in the many-body basis. While our analysis is performed in real space, the observation of intermediate fractal dimensions $0<D_2<1$ in the strong disorder regime across both approaches provides a strong indication of a multifractal regime characterized by extended yet nonergodic eigenstates.

\section{Model and Methods}
We study the isotropic random-exchange Heisenberg spin chains in one dimension with Hamiltonian
\begin{equation}
\hat{H}=\sum_{i=1}^NJ_i\hat{\bm{S}}_i\cdot\hat{\bm{S}}_{i+1}\,, \quad\quad J_i=J+b_i\,,
\label{ham}
\end{equation}
with random disorder $b_i\in[-b,b]$ uniformly distributed, and with disorder strength $b\geq0$.
In the following, we fix the value of $J$ to be $1$.
We study this system for spin lengths $S=1/2$ and $S=1$, considering various finite sizes with periodic boundary conditions, $\hat{\bm{S}}_{i+N}=\hat{\bm{S}}_i$.
Both the total spin $\hat{\bm{S}}_{\rm tot}=\sum_{i=1}^N\hat{\bm{S}}_i$ and the $z$ component of the total spin $\hat{\bm{S}}_{\rm tot}^z=\sum_{i=1}^N\hat{\bm{S}}_i^z$ commute with the Hamiltonian, i.e.,
\begin{equation}
\left[\hat{\bm{S}}_{\rm tot},\hat{H}\right]=0\,,
\quad\quad
\left[\hat{\bm{S}}_{\rm tot}^z,\hat{H}\right]=0\,.
\label{totspin}
\end{equation}
The eigenstates of the system are labeled by three quantum numbers: the energy $E$, the total spin $\hat{\bm{S}}_{\rm tot}$, and its projection along the z-axis $\hat{\bm{S}}_{\rm tot}^z$, which satisfies the following equations:
\begin{eqnarray}
\hat{H}|E,S_{\rm tot},S^z_{\rm tot}\rangle & = & E|E,S_{\rm tot},S^z_{\rm tot}\rangle\,,\\
\hat{\bm{S}}^2_{\rm tot}|E,S_{\rm tot},S^z_{\rm tot}\rangle & = & S_{\rm tot}(S_{\rm tot}+1)|E,S_{\rm tot},S^z_{\rm tot}\rangle\,,\\
\hat{S}^z_{\rm tot}|E,S_{\rm tot},S^z_{\rm tot}\rangle & = & S^z_{\rm tot}|E,S_{\rm tot},S^z_{\rm tot}\rangle\,,
\end{eqnarray}
where $S_{\rm tot}=0,1,2,...,NS$ for chains with integer total spin  and $S_{\rm tot}=1/2,3/2,....,NS$ for chains off half-integer total spin.

Next, we construct the spin-spin correlation matrix
\begin{equation}
\rho_{ij}(E,S_{\rm tot})= \langle E,S_{\rm tot},S^z_{\rm tot}| \hat{\bm{S}}_i\cdot\hat{\bm{S}}_j| E,S_{\rm tot},S^z_{\rm tot}\rangle\,,
\label{sscorr}
\end{equation}
whose elements $\rho_{ij}$ measure the correlation between the spin at site $i$ and the spin at site $j$ in a state of fixed $E$, $S_{\rm tot}$ and arbitrary $S^z_{\rm tot}$.
This quantity is independent of the quantum number $S^z_{\rm tot}$ since
\begin{equation}
\left[\hat{\bm{S}}_{\rm tot},\hat{\bm{S}}_i\hat{\bm{S}}_j\right]=0\,,
\end{equation}
and therefore
\begin{eqnarray}
& & \langle E,S_{\rm tot},S^z_{\rm tot}\pm 1| \hat{\bm{S}}_i\cdot \bm{\hat{S}}_j| E,S_{\rm tot},S^z_{\rm tot}\pm 1\rangle \nonumber\\
& & \quad = \frac{\langle E,S_{\rm tot},S^z_{\rm tot}| \hat{S}^\mp_{\rm tot}(\bm{\hat{S}}_i\cdot\bm{\hat{S}}_j)\hat{S}^\pm_{\rm tot} |E,S_{\rm tot},S^z_{\rm tot}\rangle}{S_{\rm tot}(S_{\rm tot}+1)-S^z_{\rm tot}(S^z_{\rm tot}\pm 1)} \nonumber\\
& & \quad = \frac{\langle E,S_{\rm tot},S^z_{\rm tot}| (\bm{\hat{S}}_i\cdot\hat{\bm{S}}_j)\hat{S}^\mp_{\rm tot}\hat{S}^\pm_{\rm tot} | E,S_{\rm tot},S^z_{\rm tot}\rangle} {S_{\rm tot}(S_{\rm tot}+1)-S^z_{\rm tot}(S^z_{\rm tot}\pm 1)} \nonumber\\
& & \quad = \langle E,S_{\rm tot},S^z_{\rm tot}| \bm{\hat{S}}_i\cdot\bm{\hat{S}}_j|E,S_{\rm tot},S^z_{\rm tot}\rangle\,,.
\label{equality for spin-spin correlation across multiplets}
\end{eqnarray}
Eq.~(\ref{equality for spin-spin correlation across multiplets}) guarantees that it is sufficient to work with one representative for each spin multiplet, thereby justifying the restriction to an exact-diagonalization of the subspace with lowest $S^z_{\rm tot}$.
Two sum rules naturally emerge for the correlation matrix (\ref{sscorr}):
\begin{eqnarray}
{\rm tr}\rho & = & \sum_{i=1}^N\rho_{ii}=NS(S+1) ,
\label{sumrule1}\\
\sum_{i,j=1}^N\rho_{ij} & = & \langle E,S_{\rm tot},S^z_{\rm tot}| \hat{\bm{S}}_{\rm tot}\cdot\hat{\bm{S}}_{\rm tot}|E,S_{\rm tot},S^z_{\rm tot}\rangle \nonumber\\
& = & S_{\rm tot}(S_{\rm tot}+1)\,.
\label{sumrule2}
\end{eqnarray}

The spin-spin correlator $C_r=\langle \hat{\bm{S}}_i\cdot\hat{\bm{S}}_{i+r}\rangle$ in the middle of the spectrum represent a random variable following a probability distribution $\mathcal{P}(C_r)$ that due to the translational invariance of the disorder average is independent of $i$.
We numerically investigate the decay of the spin-spin correlators with distance $r$ by calculating the average modulus of this correlator
\begin{eqnarray}
\left \langle |C_r| \right \rangle_{n,Q}= \frac{1}{Q n N} \sum_{\ell=1}^{Q} \sum_{k=1}^{n} \sum_{i=0}^{N-1}|\rho^{\ell,k}_{i,(i+r)\bmod N}| \quad,
\end{eqnarray}
where 
\begin{eqnarray}
\rho^{\ell,k}_{i,j}=\langle \psi_k^{(\ell)}|\hat{\bm{S}}_i\cdot \hat{\bm{S}}_j|\psi_k^{(\ell)}\rangle\,, \quad i,j=1,2,...,N\,, 
\label{sscorr1}
\end{eqnarray}
and where $\psi_k^{(\ell)}$ denotes the $k$-th eigenstate of the Hamiltonian in Eq.~(\ref{ham}) for the $\ell$-th disorder realization.
Here and in the following, $n$ is summed over the middle $20\%$ of the spectrum when taking averages for mid-spectrum quantities.
Note that $\psi_k^{(\ell)}$ here is the same eigenstate introduced in Eq.~(\ref{sscorr}); we have simply relabeled it for notational clarity.

Since $\rho_{ij}$ is real and symmetric, we have 
\begin{equation}
\rho | \phi _\alpha \rangle =n_\alpha | \phi _\alpha \rangle ,
\end{equation}
with real eigenvalues $n_\alpha$ and ``natural orbits''~\cite{beraManyBodyLocalizationCharacterized2015,linManybodyLocalizationSpinless2018} $|\phi _\alpha \rangle$ fulfilling the orthogonality relations
\begin{equation}
\langle \phi_\alpha|\phi_{\alpha^\prime}\rangle = \delta_{\alpha, \alpha^\prime} .
\end{equation}
We calculate the inverse participation ratio (IPR) of the natural orbitals, which is defined as~\cite{beraManyBodyLocalizationCharacterized2015,linManybodyLocalizationSpinless2018}
\begin{align}
{\rm IPR}=\frac{1}{N}\sum_{i,\alpha=1}^N |\phi_\alpha(i)|^4\,.\label{defmun}
\end{align}
The disorder average of the quantity in Eq.~(\ref{defmun}) follows
\begin{equation}
\left\langle{\rm{IPR}}\right\rangle_{\rm dis} =N\int_0^1\,dxp(x)x^2\,,
\end{equation}
where
\begin{equation}
p(x) =\left\langle \frac{1}{N^2}\sum_{\alpha=1}^N\sum_{i=1}^N \delta\left(x-|\phi_\alpha(i)|^2\right) \right\rangle_{\rm dis}\,,
\end{equation}
is the probability density for the square moduli of the components of the eigenvectors $\phi_\alpha(i)$ obtained via disorder sampling.
Note that normalization restricts the latter quantities since
\begin{equation}
\frac{1}{N}\sum_{\alpha=1}^N\sum_{i=1}^N |\phi_\alpha(i)|^2=1\,.
\end{equation}

\section{results}
\subsection{Spin-$1/2$ system}

\begin{figure}[h]
\centering
\includegraphics[width=0.48\textwidth]{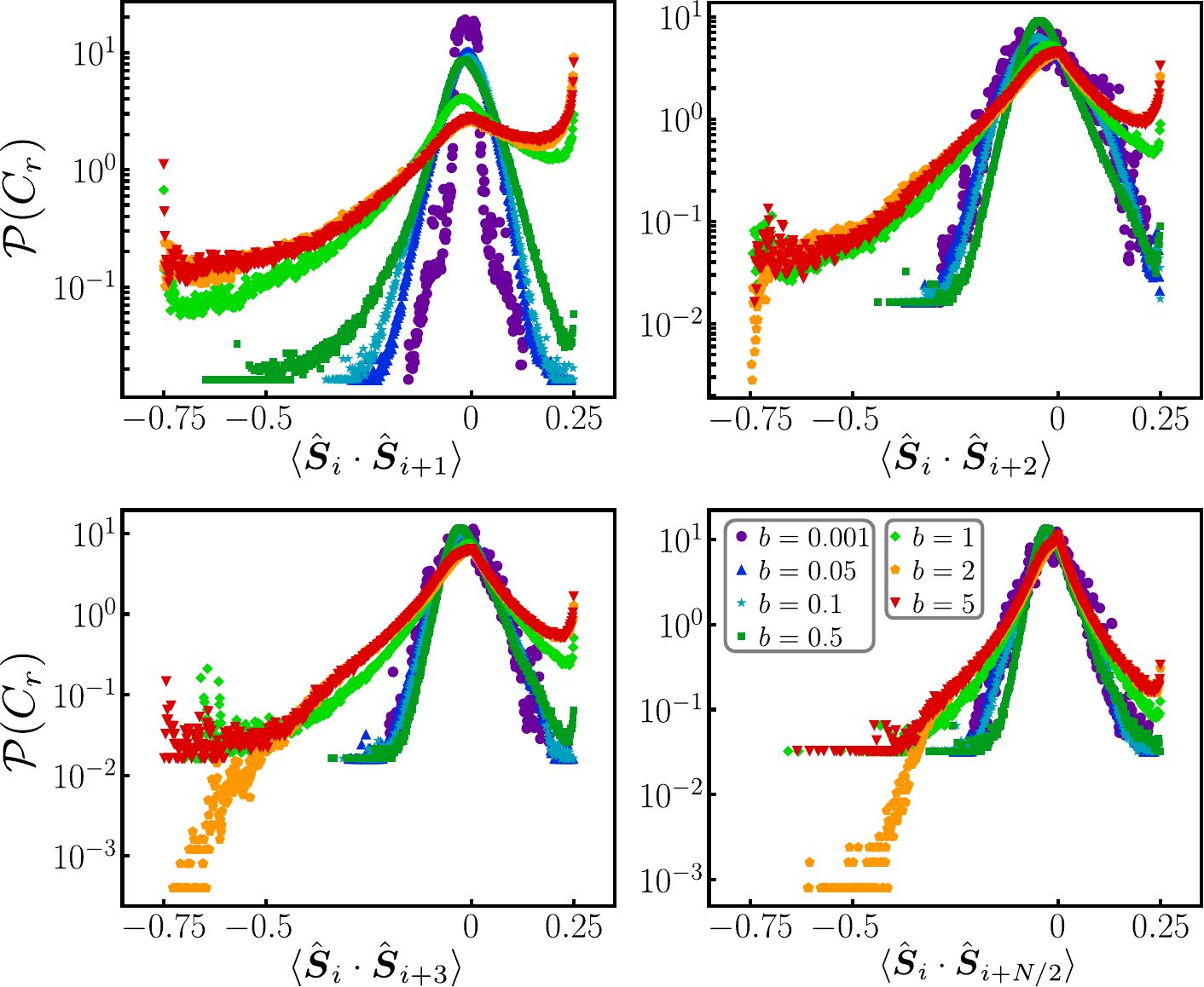}
\caption{
    \textbf{Probability distribution of spin-spin correlations $\langle \hat{\bm{S}}_i\cdot\hat{\bm{S}}_{i+r}\rangle$~\cite{pietracaprinaShiftinvertDiagonalizationLarge2018}.}
    We consider length $N=16$, distances $r=1,2,3$, and $N/2=8$, and disorder strengths $b=0.001$ to $b=5.0$.
}
\label{fig:prob_corr_N16}
\end{figure}
Fig.~\ref{fig:prob_corr_N16} shows the probability distribution $\mathcal{P}(C_r)$ of the spin-spin correlators $C_r$ for the  $S_z^{\rm tot}=0$ sector of an exchange-disordered $S=1/2$ Heisenberg chain with $N=16$ and for varying strengths of the disorder $b$.
For a very weak disorder $b=0.001$, the probability distribution of the correlators produces a sharp peak centered at zero, indicating ergodic mixing.
In this regime, the many-body eigenstates uniformly sample all microscopic configurations, so local two-spin correlations $\langle \hat{\bm{S}}_i\cdot\hat{\bm{S}}_{i+r}\rangle$ cluster around zero~\cite{nandkishoreManyBodyLocalizationThermalization2015}.
As the disorder strength increases, the distribution broadens.
In the strong‐disorder limit ($b\geq 1$), a significant fraction of nearest‐neighbor pairs $\langle \hat{\bm{S}}_i\cdot\hat{\bm{S}}_{i+1}\rangle$ behaves as independent two‐spin systems.
These eigenstates localize into nearly pure singlet or triplet configurations, where $\langle \hat{\bm{S}}_i\cdot\hat{\bm{S}}_{i+1}\rangle$ approaches the exact eigenvalues of $-3/4$ (singlet) and $1/4$ (triplet) for an isolated pair.
Consequently, the probability distribution function (PDF) of the correlators shows two sharp peaks around $-0.75$ and $0.25$, directly reflecting the dominance of local singlet and triplet bond states in the many‐body spectrum.
At $r=2$ and $r=3$, the peak around $-0.75$ becomes less pronounced but remains visible, indicating a reduced probability of singlet formation between more distant spins.
At the maximal separation $r=N/2=8$, even strong disorder cannot sustain true long-range singlet formation: the PDF remains predominantly centered around zero, with only faint shoulders at $-0.75$ suggesting rare instances of distant spin-singlet states.
Notably, the peak around $0.25$ remains nearly as prominent up to $r=N/2$, highlighting the asymmetry in triplet versus singlet correlations at large distances.
The origin of this asymmetry as an artifact due to the $SU(2)$ symmetry of the model is discussed in \cref{appendix unique eigenstate}.

\begin{figure}[h]
\centering
\includegraphics[width=0.35\textwidth]{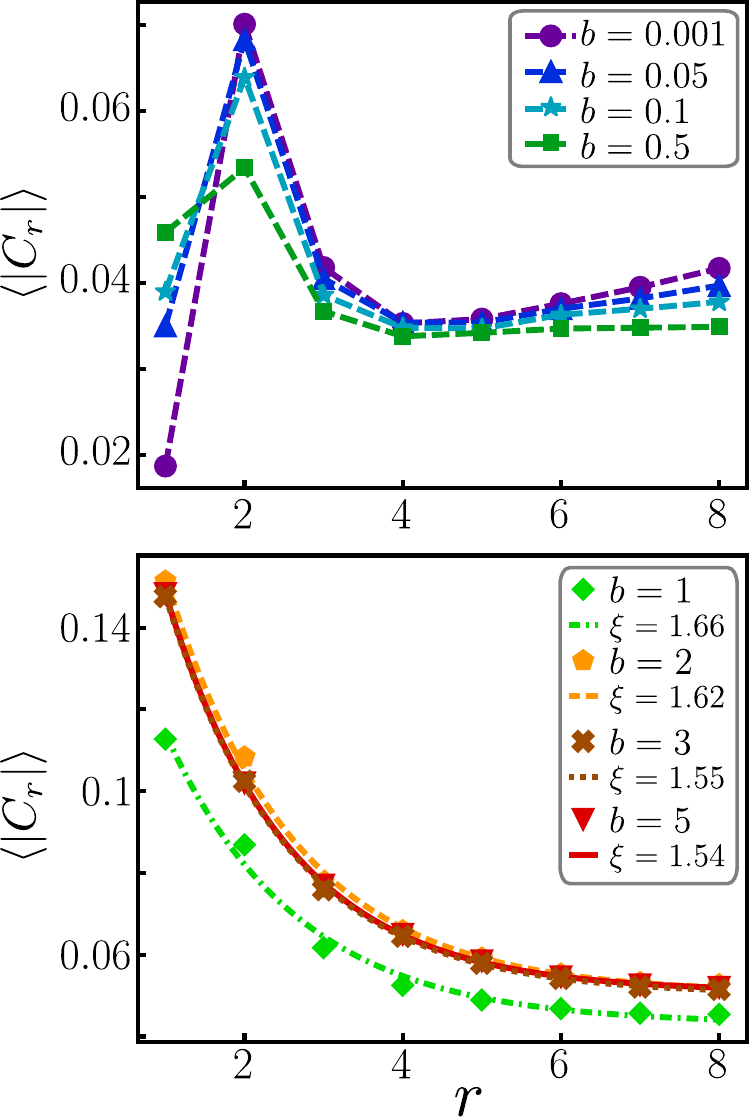}
\caption{
    \textbf{Average modulus of correlation $\langle |\hat{\bm{S}}_i\cdot\hat{\bm{S}}_{i+r}| \rangle $ as a function of distance $r$.}
    We consider length $N=16$, distances $r=1,2,3$, and $N/2=8$, and disorder strengths $b=0.001$ to $b=5.0$. We fit $\left \langle |C_r| \right \rangle \approx Ae^{-r/\xi}+ B$ for each disorder strength.
}
\label{fig:corr_r_N16}
\end{figure}

Fig.~\ref{fig:corr_r_N16} shows the decay of the average modulus of the mid‐spectrum correlators with distance.
In the weak disorder regime $b=0.001,0.05, 0.1, 0.5$, $\left \langle |C_r|\right \rangle \approx (0.02,0.07)$ remains essentially flat as a function of $r$, which is consistent with an effective infinite-temperature ensemble and delocalized spins. 
In contrast, for strong disorder strength $b=1.0,2.0,3.0,5.0$, $\left \langle |C_r| \right \rangle \approx Ae^{-r/\xi}+ B$ decays exponentially with fitted correlation lengths $\xi\approx 1.5-1.7$.
Physically, $\xi\approx 1.5-1.7$ indicates that each spin is essentially locked into a nearest-neighbor singlet or triplet, so correlation beyond one or two sites is exponentially suppressed.

Next, we compute the inverse participation ratio (IPR) as defined in Eq.~(\ref{defmun}) and average it over $Q$ disorder realizations.
To do this, we construct the $N \times N$ spin-spin correlation matrix as defined in Eq.~(\ref{sscorr1}) for each mid‐spectrum eigenstate $|\psi\rangle$ and all disorder realizations.
The eigenvectors correspond to "modes" describing the distribution of the two-spin correlations across the chain.
When the disorder is weak $b<<1$, the eigenvectors $\phi_\alpha$ are broadly “delocalized”, so their IPRs scale as $\mathcal{O}(1/N)$.
Correspondingly, the distribution of components $\{|\phi_\alpha(i)|^2\}$ over all $i$ should be equal.
As disorder increases into the random‐singlet or MBL‐like regime for $b\geq 1$, some modes collapse onto one or two sites.
Since a perfect singlet shares equal weight on two adjacent sites $i$ and $i+1$, this gives $|\phi(i)|^2=|\phi(i+1)|^2=1/2$.

\begin{figure}[ht]
\centering
\includegraphics[width=0.40\textwidth]{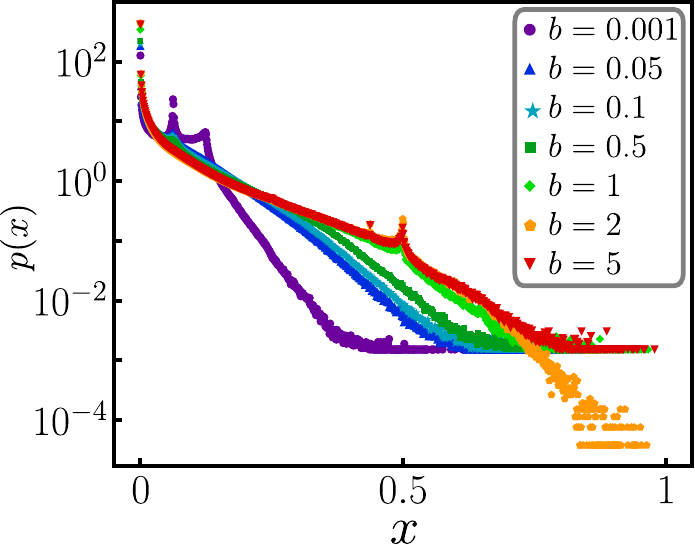}
\caption{
    \textbf{Probability density $p(x)$ of the average squared correlation‐matrix eigenvector components $|\phi_k(i)|^2$.}
    The average runs over all modes $k=1,2,...,N$, sites $i=1,2,...,N$, the eigenstates in the middle $20\%$ of the spectrum, and $Q=500$ disorder realizations for length $N=16$.
    Curves are shown for disorder amplitudes $b=0.001, 0.1, 0.5, 1.0, 2.0, 5.0$.
    }\label{fig:prob_N16}
\end{figure}

Fig.~\ref{fig:prob_N16} shows the probability density $p(x)$ of the average modulus of the squared eigenvector components $|\phi_\alpha(i)|^2$ for the correlation matrices $\rho^{\ell,k}$ for $N=16$ and $Q=500$ disorder realizations. 
For small disorder strengths $b=0.001-0.1$, $p(x)$ exhibits a sharp peak near $x=1/N\approx 0.0625$, signaling that each mode is delocalized over all $N$ sites as expected in the ETH regime.
This is consistent with our earlier observation that, for these weak disorders, the two‐point correlator histograms are sharply centered at $C_r=0$ for all $r$, reflecting infinite-temperature random‐matrix fluctuations, and that the average modulus of the correlator $\langle |C_r|\rangle$ is nearly flat as a function of $r$.
As $b$ increases to $0.5$, the dominant $x=1/N$ peak broadens, indicating that moderate disorder has induces a clustering of correlation eigenmodes on a smaller subset of sites, thereby raising their inverse participation ratios (IPRs) above $1/N$.
However, no sharply localized two-site singlets have yet formed in this regime.
In the strong-disorder regime $b\geq 1$, a pronounced delta-like peak appears at $x=0.5$.
These modes correspond to near-perfect two-site singlets or triplets (each frozen on two adjacent sites): each such eigenvector has $|\phi(i)|^2=|\phi(i+1)|^2$ and vanishes on all other sites.
Since the evaluation of the probability density runs over $N$ modes and $N$ sites per mode, each frozen singlet contributes exactly $N-2$ times with $x=0$ and twice with $x=1/2$.
Hence, $p(x)$ also develops a pronounced peak at $x=0$ (the log-scale “zero peak”) arising from those exact zeros.
Physically, once nearest-neighbor bonds “freeze” into singlets or triplets, as seen in the $\mathcal{P}(C_r)$ distribution, the corresponding correlation eigenmode is locked onto precisely two sites with $|\phi(i)|^2=|\phi(i+1)|^2$.

\begin{figure}[ht]
\centering
\includegraphics[width=0.40\textwidth]{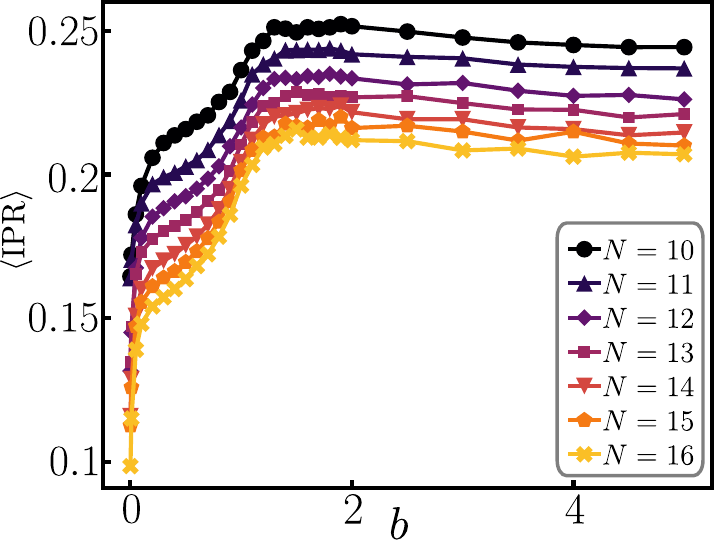}
\caption{
    \textbf{Disorder‐averaged inverse participation ratio ${\rm IPR}=\frac{1}{N}\sum |\phi_k^{q,\alpha}(i)|^4$ of the spin-spin correlation‐matrix eigenmodes.}
    Shown is the IPR as a function of disorder strength $b$ for chain lengths $N=10-16$ and $Q=500$ disorder realizations.
}
\label{fig:IPR_S05}
\end{figure}
Next, we turn to the inverse‐participation ratio (IPR) of the correlation‐matrix eigenmodes to quantify the increasing localization of the modes with increasing disorder.
In Fig.~\ref{fig:IPR_S05}, we plot $\langle {\rm IPR \rangle}$ versus $b$ for chain lengths $N=10-16$ with spin length $S=1/2$.
In the weak disorder regime $b<1$, the average value of the IPR falls with increasing $N$, tracking the ergodic scaling $\sim 1/N$.
In contrast, for the strong disorder regime $b\geq 1$, all curves collapse onto a common plateau around $0.21-0.25$.
This intermediate value, well above $1/N$ but below the ideal two‐site singlet limit $1/2$, signifies that a finite fraction of modes have “frozen” into almost perfect nearest‐neighbor singlets or triplets, each contributing IPR$\sim 1/2$, while the remainder retains some residual weight spread over more sites.

\begin{figure}[ht]
\centering
\includegraphics[width=0.40\textwidth]{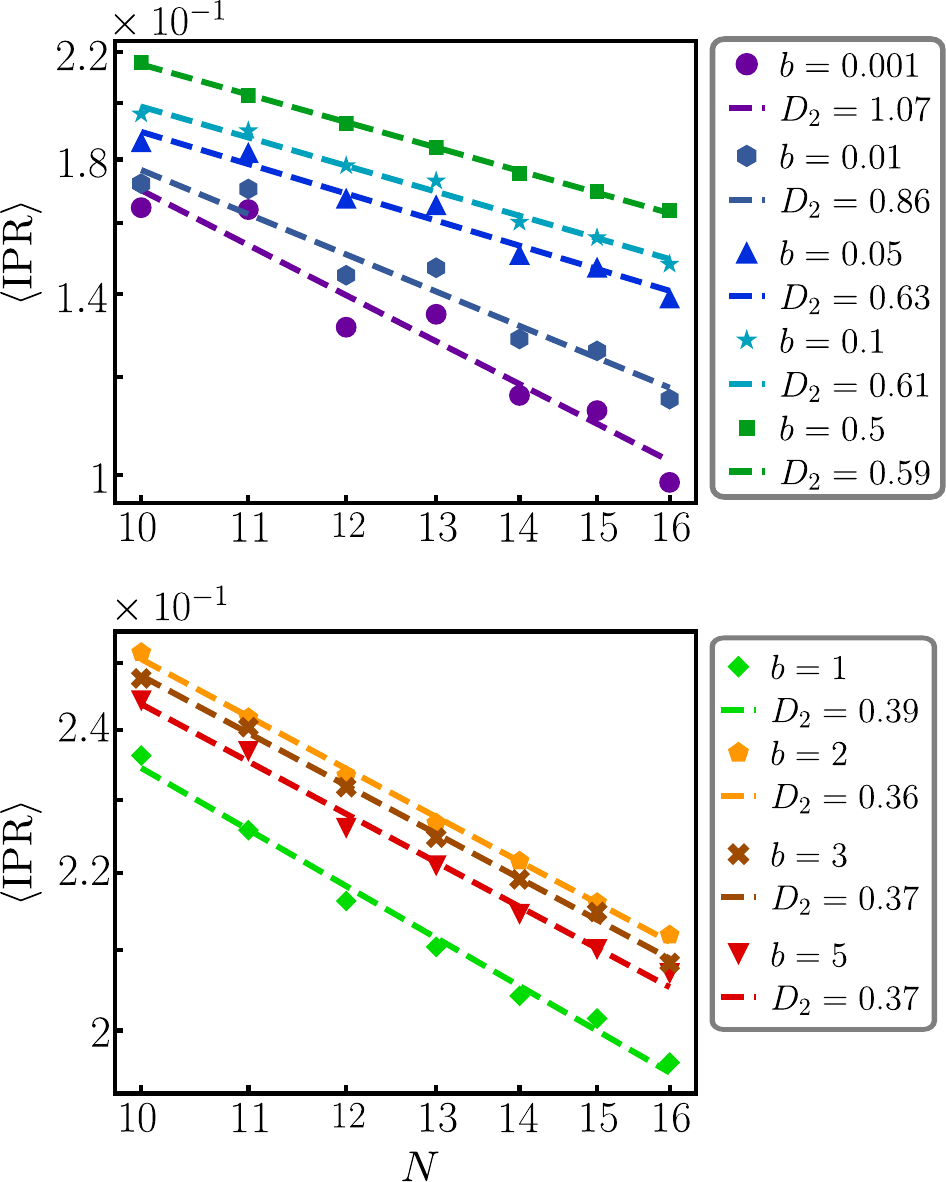}
\caption{
    \textbf{Scaling of the IPR as a function of system size $N$.}
    Shown is the IPR as function of $N$ in log-log scale for different disorder strengths $b$.
    We include a fit to the IPR as IRP$\sim N^{-D_2}$ to extract the correlation-dimensional exponent.
}
\label{fig:IPR_scaling_S05}
\end{figure}
In order to quantify how the spatial support of the correlation‐matrix modes evolves with system size, we have performed a finite‐size scaling of the normalized IPR for several disorder strengths with fitting $\langle {\rm IPR} \rangle \sim N^{-D_2}$, where $D_2$ is the so‐called “correlation‐dimensional” exponent (the fractal dimension of the modes).
The upper panel of Fig.~\ref{fig:IPR_scaling_S05} shows the weak‐to‐intermediate disorder regime $b=0.001, 0.01,0.05,0.1,0.5$ on a log-log plot of the IPR plotted against $N$.
For very weak disorder $b=0.001$, we find $D_2\approx 1.04$, i.e., $\langle {\rm IPR} \rangle \sim 1/N$, confirming a uniform spread of all correlation modes over the chain, consistent with an ergodic phase.
As disorder increases to $b=0.0.01,0.05,0.1,0.5$, the exponent drops to $D_2 \approx 0.6-0.86$, signaling that eigenmodes become partially localized or “multifractal”, where each mode occupies only a sub‐extensive fraction $N^{D_2}$ of sites.
In the strong disorder regime $b\geq1.0$, the exponent further decreases to $D_2\approx 0.36-0.39$, but does not reach zero.
A nonzero $D_2$ for large disorder shows that no localized phase is reached, in line with earlier observations~\cite{sieglImperfectManyBodyLocalization2023,protopopovNonAbelianSymmetriesDisorder2020,protopopovEffectSU2Symmetry2017}.
Instead, there remain modes that spread over a sub-extensive number $N^{D_2}$ of sites.

\subsection{Spin$-1$ system}
We extend our analysis of spin-spin correlations to the spin-1 Heisenberg chain in Eq.~(\ref{ham}).
We again work with the middle $20\%$ of the spectrum.
Again, we start by studying the probability distribution of the spin-spin correlator $\mathcal{P}(C_r)$ for the system size $N=10$.
\begin{figure}[h]
\centering
\includegraphics[width=0.48\textwidth]{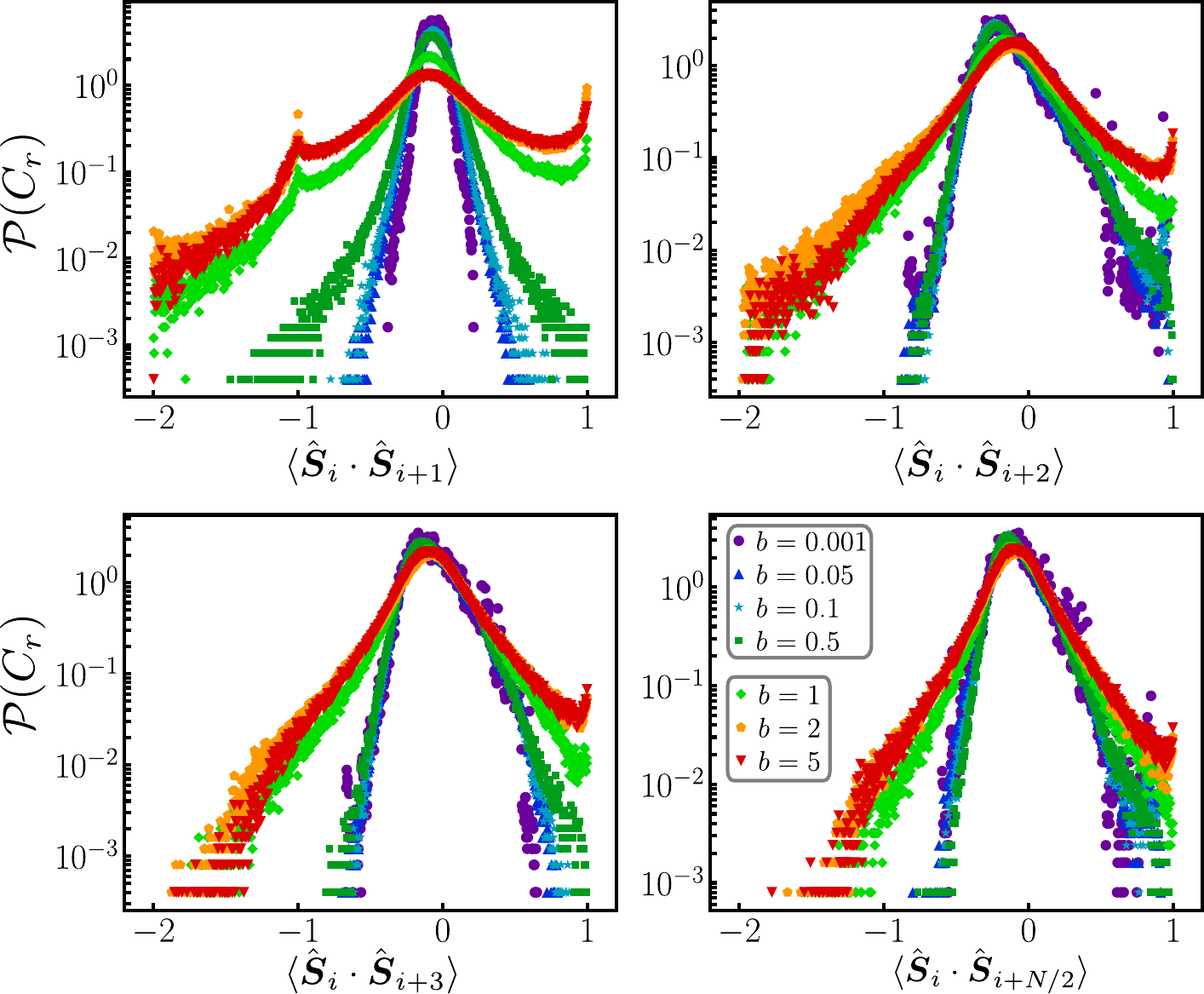}
\caption{
    \textbf{Probability distribution of spin-spin correlations $\langle \hat{\bm{S}}_i.\hat{\bm{S}}_{i+r} \rangle $ for $S=1$}.
    We considered the $S_z^{\rm tot}=0$ sector of the exchange-disordered $S=1$ Heisenberg chain with length $N=10$, distances $r=1,2,3$, and $N/2=5$ with disorder strengths $b=0.001$ to $b=5.0$.
}\label{fig:prob_corr_N10_S1}
\end{figure}
At weak disorder, $\mathcal{P}(C_r)$ peaks near zero for every $r$ as shown in Fig.~\ref{fig:prob_corr_N10_S1}, consistent with ergodic behavior where eigenstates uniformly sample the Hilbert space.
As disorder increases $b\geq 1$, $\mathcal{P}(C_1)$ develops sharp peaks around $-2$ (singlet), $-1$ (triplet), and $+1$ (quintet), signaling the emergence of localized dimer states.
A weaker shoulder near $-2$ reflects the singlet’s distinct stability compared to higher-spin multiplets.
These correlations decay rapidly with distance: for $r=2$ and $r=3$, the peaks broaden and diminish.
At $r=5$, even under the largest $b$, the PDF is overwhelmingly peaked around zero, demonstrating that long-range dimerization is negligible for spin$-1$ chain as well.
The coexistence of singlet, triplet, and quintet dimer states distinguishes spin-1 systems from their spin-1/2 counterparts, where only $C_1=-3/4$ or $1/4$ occurs, and highlights the role of local Hilbert space geometry in disorder-induced localization.

\begin{figure}[h]
\centering
\includegraphics[width=0.35\textwidth]{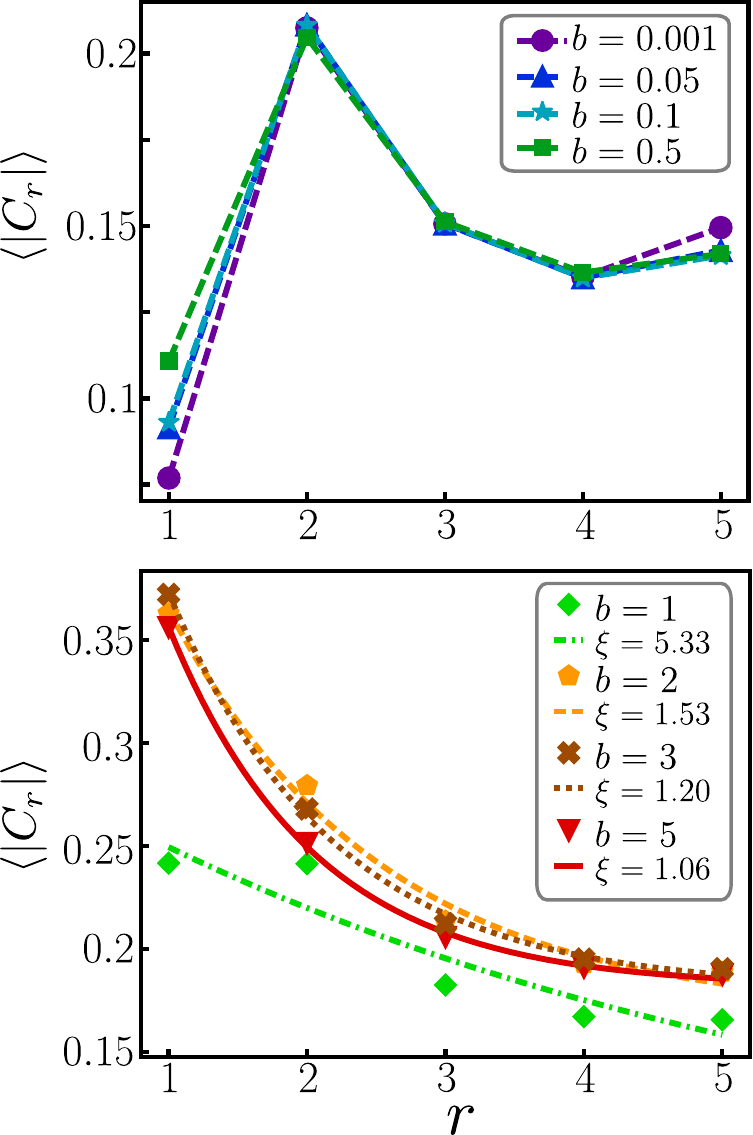}
\caption{
    \textbf{Average modulus of the correlator $\langle |\mb{S}_i.\mb{S}_{i+r}| \rangle $ as a function of distance $r$.}
    We considered the $S_z^{\rm tot}=0$ sector of the exchange-disordered $S=1$ Heisenberg chain with length $N=10$, distances $r=1,2,3$, and $N/2=5$ with disorder strengths $b=0.001$ to $b=5.0$.
}\label{fig:abs_corr_N10_S1}
\end{figure}
In Fig.~\ref{fig:abs_corr_N10_S1}, we show the average modulus of the correlator $\langle |C(r)| \rangle$.
At weak disorder, $\langle |C(r)| \rangle$ exhibits a nearly $r$-independent plateau ($\approx 0.08-0.2$), characteristic of ergodicity.
In contrast, for $b \geq 1$ the modulus of the correlator transitions to an exponential decay with increasing $r$, thereby signaling the onset of localization.
The correlation length at $b=1.0$ is $\xi \approx 5.3$, notably longer than in the spin-$1/2$ case.
This increase of the correlation length is a consequence of the larger local Hilbert space in the spin-1 system.
As disorder strengthens further $b=5.0$, $\xi$ shrinks to $\approx 1.1$, confirming that correlations are sharply confined to nearest neighbors.

\begin{figure}[h]
\centering
\includegraphics[width=0.40\textwidth]{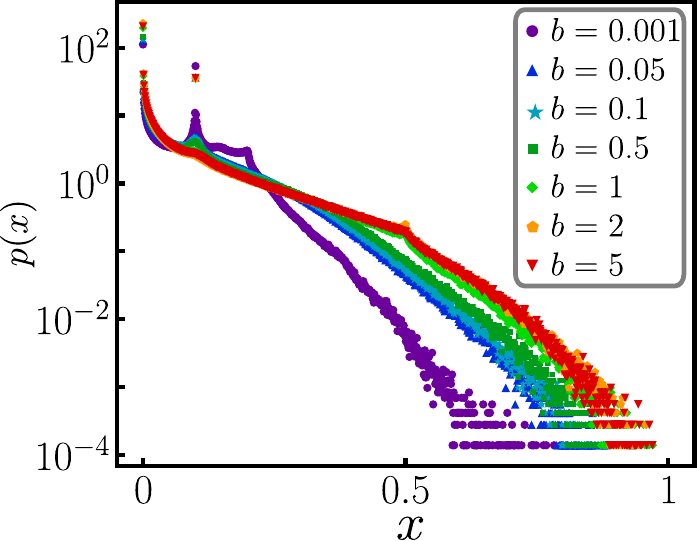}
\caption{
    \textbf{Probability density $p(x)$ of squared correlation‐matrix eigenvector components
            $|\phi_k(i)|^2$.}
    We consider all modes $k=1,2,...,N$, sites $i=1,2,...,N$, and $Q=500$ disorder realizations for a chain length of $N=10$.
    The curves represent disorder amplitudes $b=0.001, 0.1, 0.5, 1.0, 2.0, 5.0$.
    }\label{fig:x_dis_N10_S1}
\end{figure}
Fig.~\ref{fig:x_dis_N10_S1} shows the distribution $p(x)$ of the on-site squared amplitudes of the eigenmodes of the correlation matrix.
In the weak-disorder regime, $p(x)$ is peaked at $x\approx 1/N=0.1$, reflecting fully extended states.
Upon increasing $b$ into the strong-disorder regime, $p(x)$ develops a peak around $x=0.5$, but less prominent as compared to the spin-$1/2$ counterparts.
This suppression reflects fundamental differences in how localization manifests in higher-spin systems.
For spin-1 dimers, only the two most strongly localized eigenstates, typically singlet and triplet configurations, contribute to this peak, while the remaining seven eigenstates, including higher-spin quintets and more extended configurations remain delocalized.
This behavior contrasts with the spin-$1/2$ case, where half of the states (2 out of 4) can localize onto a single bond.
This dilution effect emerges because the larger local Hilbert space of spin-1 systems provides more pathways for quantum fluctuations to resist complete localization, even under strong disorder.
The weaker peak at $x=0.5$ thus directly reflects the competition between disorder-induced localization and the enhanced quantum-mechanical "mixing capacity" inherent to higher-spin systems.

\begin{figure}[h]
\centering
\includegraphics[width=0.40\textwidth]{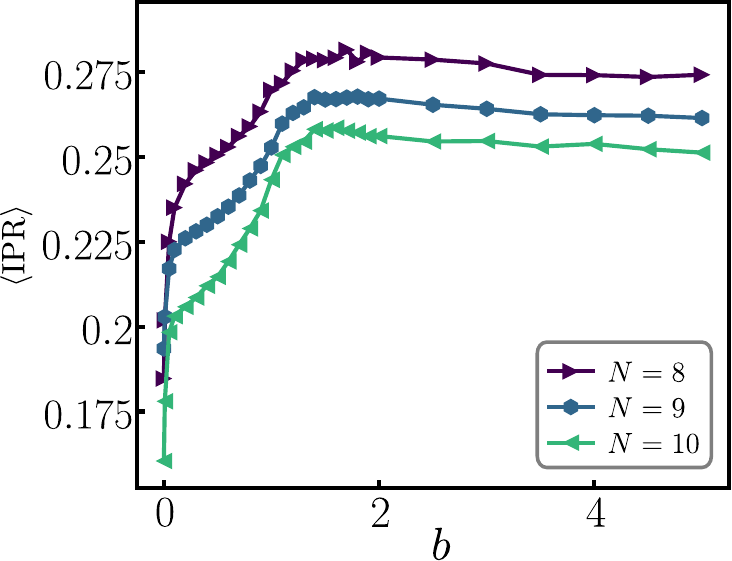}
\caption{
    \textbf{Disorder‐averaged inverse participation ratio ${\rm IPR}=\frac{1}{N}\sum |\phi_k^{q,\alpha}(i)|^4$ of the spin-spin correlation‐matrix eigenmodes.},
    We show the IPR as a function of disorder strength $b$ for chain lengths $N=8-10$ and averaged over $Q=500$ disorder realizations.
}\label{fig:IPR_S1}
\end{figure}
Fig.~\ref{fig:IPR_S1} shows the average IPR $\langle \rm{IPR} \rangle$ as a function of the disorder strength $b$ for spin-1 chains of length $N=8-10$.
In the ergodic regime, $\langle \rm{IPR} \rangle$ again decreases with $N$, consistent with extended modes.
Once $b\geq 1$, $\langle \rm{IPR} \rangle$ saturates to values $\approx 0.25-0.28$, intermediate between the ergodic ($\sim 1/N$) and fully localized ($\sim 0.5$) limits.
This plateau indicates that only a fraction of the total modes can fully localize onto two sites, while the rest remains partially delocalized.

\begin{figure}[ht]
\centering
\includegraphics[width=0.40\textwidth]{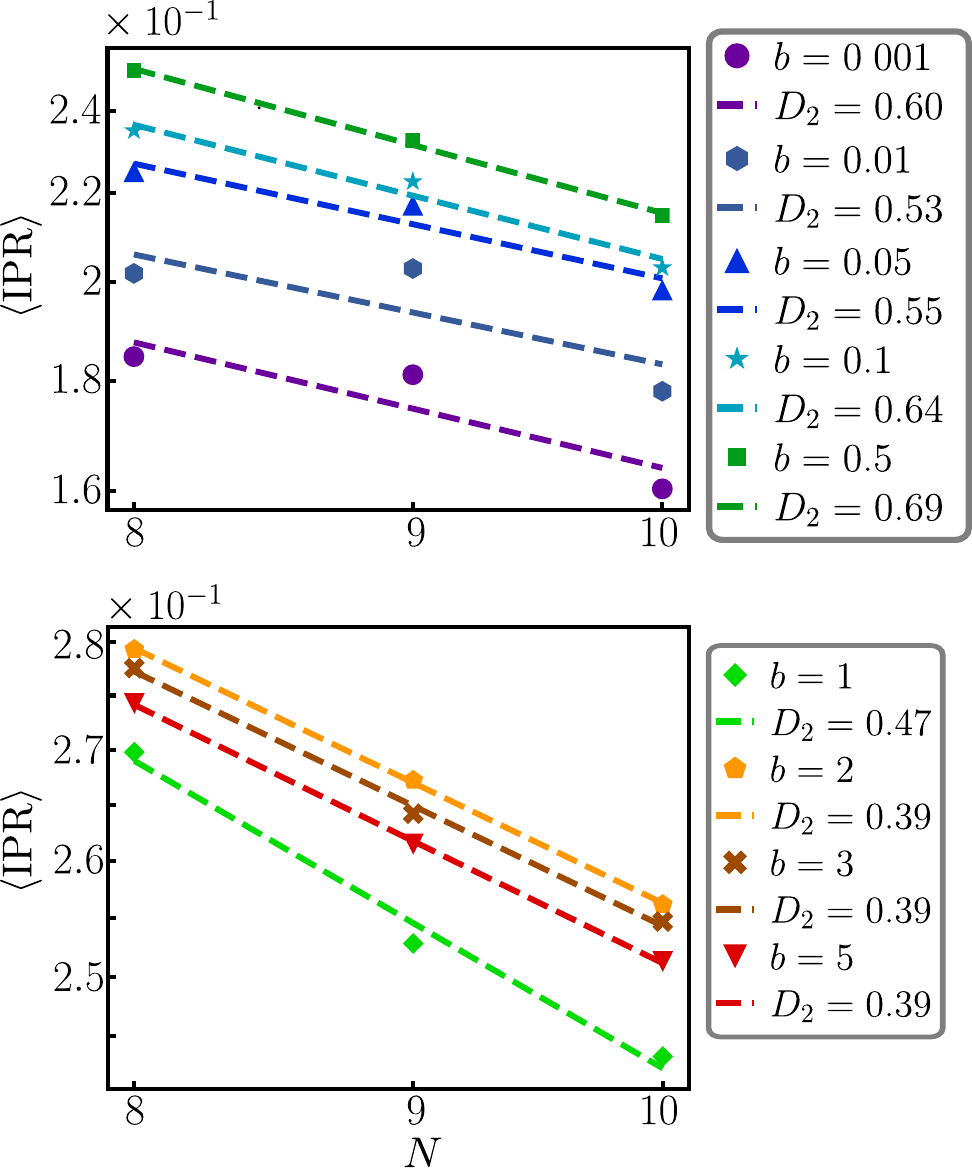}
\caption{
    \textbf{Scaling of the IPR as a function of system size $N$}
    Shown is the IPR plotted against the chain length $N$ in a log-log scale.
    We added a fit to the IPR as IPR$\sim N^{-D_2}$ in order to extract the correlation-dimensional exponents $D_2$.
}\label{fig:IPR_scaling_S1}
\end{figure}
Finally, we show the fractal dimension analysis in Fig.~\ref{fig:IPR_scaling_S1}.
The (upper panel) presents the weak-to-intermediate disorder regime: even at $b=0.001$ the fitted $D_2=0.6$ is well below unity.
The extracted fractal dimension $D_2$, obtained from finite-size scaling with $N=8,9,10$, deviates from the expected ergodic value $1$, possibly due to finite-size effects and limited data points.
We expect that for larger $N$, one will recover $D_2 \to 1$ in the $b\to 0$ limit.
The lower panel of Fig.~\ref{fig:IPR_scaling_S1} shows the strong disorder limit, where $D_2$ drops to $\approx 0.39-0.47$.
This decrease is similar to the spin-$1/2$ plateau, indicating that two-site localization dominates while the remaining modes retain a sub-extensive support.

\section{Conclusion and Discussion}
In conclusion, our comprehensive analysis of two-point spin correlations in $1D$ exchange-disordered Heisenberg spin chains with common spin lengths $S=1/2$ and $S=1$ across varying disorder strengths reveals a clear crossover from an ergodic ETH regime to a quasi-localized multifractal phase.
Through analysis of mid-spectrum eigenstates, we analyzed the full distribution of spin correlators and their spatial decay.
We diagonalized the spin-spin correlation matrix to study its eigenmodes, including eigenvector probability profiles and IPR scaling.
In the weak-disorder, spin correlations exhibit a narrow distribution around $C_r=0$, while correlation-matrix eigenvectors delocalize across the system, with IPR $\sim 1/N$, consistent with thermal behavior.
In stark contrast, strong disorder drives the system into a quasi-localized multifractal phase marked by two key features: first, the correlator distribution develops peaks at certain values corresponding to singlet-like and triplet-like bonds with correlations whose modulus decays exponentially with distance.
Secondly, the correlation matrix yields emergent dimer-like eigenmodes with a fraction of eigenvectors concentrated on two strongly entangled spins, while others retain partial delocalization.
A finite‐size scaling unambiguously demonstrates a fractal dimension $D_2\approx 0.37-0.39$ in the strong‐disorder regime, far from both the ergodic value ($D_2=1$) and the fully localized limit ($D_2=0$), indicating a stable multifractal phase.
This intermediate $D_2$ reflects the system’s competition between full localization and $SU(2)$ symmetry.
While disorder favors dimerization, quantum fluctuations prevent complete two-site freezing, instead stabilizing a fractal ensemble of localized spin pairs.
Our results establish a unified framework for understanding how $SU(2)-$invariant spin chains evade full localization, offering broader insights into the interplay of symmetry and disorder in quantum many-body systems.
While providing a detailed static characterization of correlation spectra and their multifractal scaling, the restricted system sizes $N\leq 16$ for $S=1/2$ and $N\leq 10$ for $S=1$ in our study leave open fundamental questions about the thermodynamic stability of the observed multifractal phase.
To overcome this bottleneck, a promising avenue is to deploy machine‐learning techniques.
Neural-network classifiers trained on spin-correlator distributions could identify phase boundaries with minimal size dependence, while diffusion-map algorithms applied to IPR spectra might predict fractal dimensions in regimes inaccessible to exact diagonalization.
Such approaches have already demonstrated success in detecting hidden order parameters in frustrated magnets~\cite{carrasquillaMachineLearningPhases2017} and MBL transitions~\cite{beachMachineLearningVortices2018}.
On the dynamic side, it will be crucial to connect our static fingerprints of multifractality to real‐time behavior.
For example, tracking the spread of local spin excitations could distinguish between multifractal and localized dynamics.
Out-of-time-order correlators (OTOCs) would also test whether the multifractal regime exhibits characteristic logarithmic spreading~\cite{vasseurParticleholeSymmetryManybody2016} intermediate between thermal and MBL behavior.

\section{Acknowledgments}
We thank R. A. R\"omer and Y. Gao for discussion and for sharing results with us prior to publication. The work of J. Siegl was funded by project B09 of CRC 1277 of the Deutsche Forschungsgemeinschaft.

\section{Data Availability}
The data supporting the figures in this manuscript is available at \url{https://epub.uni-regensburg.de/77459/}.

\appendix 

\section{Special eigenvalues of the spin-spin correlation matrix\label{appendix unique eigenstate}}
As indicated by the numerics, there are special values in the spectrum of the correlation matrix.
Furthermore, such values also present themselves in the probability distribution of the IPR where they contribute sharply peaked features distinct from a smooth background.
We present here a brief explanation of their origin, identifying them as signatures of symmetry properties of the model.
An additional restriction on the spin-spin correlation matrix occurs in the lone multiplet of the maximal total spin $S_{\rm tot}=NS$.
Evaluation of the matrix in the state of maximum $S_{\rm tot}$ gives
\begin{equation}
\rho=S\,\mathbf{1}+S^2\,
\left(
\begin{array}{ccc}
1 & 1 & \cdots \\
1 & 1 & \cdots \\
\vdots & \vdots & \ddots
\end{array}
\right)\,,
\label{sscorrSmax}
\end{equation}
and an obvious eigenvector is
\begin{equation}
\bar\phi=\frac{1}{\sqrt{N}}\left(1,\dots,1\right)^T\, ,
\label{barphi}
\end{equation}
with the eigenvalue.
\begin{equation}
\bar n=S+NS^2\,.
\end{equation}
All other eigenvectors are orthogonal to $\bar{\phi}$ and share $(N-1)$-fold degenerate eigenvalue of $S$. To verify this degeneracy explicitly, we examine the second matrix appearing on the right-hand side of Eq.~(\ref{sscorrSmax}), which consists entirely of ones. Its eigenvalues are obtained by computing the characteristic polynomial. We define two auxiliary $N \times N$ matrices:
\begin{equation}
A_N(\lambda)=
\left(
\begin{array}{cccc}
1-\lambda & 1 & \cdots & 1 \\
1 & 1-\lambda & \cdots & 1\\
\vdots & \vdots & \cdots & \vdots\\
1 & 1 & \cdots & 1-\lambda
\end{array}
\right)\,,
\label{A_N}
\end{equation}
and 
\begin{equation}
B_N(\lambda)=
\left(
\begin{array}{cccc}
1 & 1 & \cdots & 1 \\
1 & 1-\lambda & \cdots & 1\\
\vdots & \vdots & \cdots & \vdots\\
1 & 1 & \cdots & 1-\lambda
\end{array}
\right)\,.
\label{B_N}
\end{equation}
We will use the method of induction to find the ${\rm det}[A_N(\lambda)]$, which is the characteristic polynomial of the second matrix in the rhs of Eq.~(\ref{sscorrSmax}). One can check for $N=2$, ${\rm det}[A_2(\lambda)]=\lambda (\lambda-2)$ and ${\rm det}[B_2(\lambda)]=-\lambda$. Let's assume that for system size $N$,
\begin{align}
{\rm det}\,[A_N(\lambda)] &= (-1)^N\,\lambda^{N-1}(\lambda - N) ,
\label{detA} \\
{\rm det}\,[B_N(\lambda)] &= -(-1)^N\,\lambda^{N-1} .
\label{detB}
\end{align}
Then we must prove that the formula holds for the system size $N+1$ as well. Now, $A_{N+1}(\lambda)$
is a matrix of dimension $(N+1) \times (N+1)$. We subtract the second row of the matrix from the first row to get
\begin{align}
 A_{N+1}(\lambda)&=
\left(
\begin{array}{ccccc}
-\lambda & \lambda & 0 & \cdots & 0 \\
1 & 1-\lambda & 1 & \cdots & 1\\
1 & 1 & 1-\lambda & \cdots & 1 \\
\vdots & \vdots & \vdots  & \cdots & \vdots\\
1 & 1 & 1& \cdots & 1-\lambda
\end{array}
\right)\, \nonumber  \\ \nonumber \\
&=-\lambda\, {\rm det}[A_N(\lambda)]-\lambda\,{\rm det}[B_N(\lambda)] \nonumber \\ \nonumber \\
&=-\lambda (-1)^N\lambda^{N-1}(\lambda-N)+\lambda(-1)^N\lambda^{N-1}\, \nonumber \\ \nonumber \\
&=(-1)^{N+1}\lambda^N(\lambda-N-1).
\label{A_Np1v2}   
\end{align}
Applying the same row operations to $B_{N+1}(\lambda)$ gives
\begin{align}
 B_{N+1}(\lambda)&=
\left(
\begin{array}{ccccc}
0 & \lambda & 0 & \cdots & 0 \\
1 & 1-\lambda & 1 & \cdots & 1\\
1 & 1 & 1-\lambda & \cdots & 1 \\
\vdots & \vdots & \vdots  & \cdots & \vdots\\
1 & 1 & 1& \cdots & 1-\lambda
\end{array}
\right)\, \nonumber  \\ \nonumber \\
&=-\lambda\, {\rm det}[B_N(\lambda)]\nonumber \\ \nonumber \\
&=-(-1)^{N+1}\lambda^N\,.
\label{B_Np1v2}   
\end{align}
Therefore, Eq.~(\ref{detA}) and Eq.~(\ref{detB}) holds for any $N$. Hence, the roots of the characteristic equation of Eq.~(\ref{detA}) are,
\begin{equation}
    \lambda=0\,({\rm multiplicity~~ N-1})\, ,\quad \quad \lambda=N\, .
\end{equation}

Moreover, the vector Eq.~(\ref{barphi}) fulfills, in accordance with Eq.~(\ref{sumrule2}), 
\begin{equation}
\bar\phi^T\rho(E,S_{\rm tot})\bar\phi
=\frac{1}{N}S_{\rm tot}(S_{\rm tot}+1)\,,
\label{sumrule3}
\end{equation}
but is in general not an eigenvector of the correlation matrix except for the case $S_{\rm tot}=0$ where we have
\begin{equation}
\rho(E,0)\bar\phi
=\frac{1}{N}\left(
\begin{array}{c}
\langle E,0,0|
\hat{\bm{S}}_1\cdot\hat{\bm{S}}_{\rm tot}|E,0,0\rangle \\
\vdots \\
\langle E,0,0|
\hat{\bm{S}}_N\cdot\hat{\bm{S}}_{\rm tot}|E,0,0\rangle 
\end{array}
\right)=0\,.
\label{sumrule4}
\end{equation}
Thus, the correlation matrix of any singlet state has an eigenvalue of zero with the corresponding eigenvector Eq.~(\ref{barphi}).
The above restrictions on the spin-spin correlation matrix arise entirely from the underlying SU(2) invariance.
This symmetry might possibly also lead to further constraints that are more subtle to detect.
\bibliography{bibliography}

\begin{thebibliography}{49}%
\makeatletter
\providecommand \@ifxundefined [1]{%
 \@ifx{#1\undefined}
}%
\providecommand \@ifnum [1]{%
 \ifnum #1\expandafter \@firstoftwo
 \else \expandafter \@secondoftwo
 \fi
}%
\providecommand \@ifx [1]{%
 \ifx #1\expandafter \@firstoftwo
 \else \expandafter \@secondoftwo
 \fi
}%
\providecommand \natexlab [1]{#1}%
\providecommand \enquote  [1]{``#1''}%
\providecommand \bibnamefont  [1]{#1}%
\providecommand \bibfnamefont [1]{#1}%
\providecommand \citenamefont [1]{#1}%
\providecommand \href@noop [0]{\@secondoftwo}%
\providecommand \href [0]{\begingroup \@sanitize@url \@href}%
\providecommand \@href[1]{\@@startlink{#1}\@@href}%
\providecommand \@@href[1]{\endgroup#1\@@endlink}%
\providecommand \@sanitize@url [0]{\catcode `\\12\catcode `\$12\catcode
  `\&12\catcode `\#12\catcode `\^12\catcode `\_12\catcode `\%12\relax}%
\providecommand \@@startlink[1]{}%
\providecommand \@@endlink[0]{}%
\providecommand \url  [0]{\begingroup\@sanitize@url \@url }%
\providecommand \@url [1]{\endgroup\@href {#1}{\urlprefix }}%
\providecommand \urlprefix  [0]{URL }%
\providecommand \Eprint [0]{\href }%
\providecommand \doibase [0]{https://doi.org/}%
\providecommand \selectlanguage [0]{\@gobble}%
\providecommand \bibinfo  [0]{\@secondoftwo}%
\providecommand \bibfield  [0]{\@secondoftwo}%
\providecommand \translation [1]{[#1]}%
\providecommand \BibitemOpen [0]{}%
\providecommand \bibitemStop [0]{}%
\providecommand \bibitemNoStop [0]{.\EOS\space}%
\providecommand \EOS [0]{\spacefactor3000\relax}%
\providecommand \BibitemShut  [1]{\csname bibitem#1\endcsname}%
\let\auto@bib@innerbib\@empty
\bibitem [{\citenamefont
  {Deutsch}(1991)}]{deutschQuantumStatisticalMechanics1991}%
  \BibitemOpen
  \bibfield  {author} {\bibinfo {author} {\bibfnamefont {J.~M.}\ \bibnamefont
  {Deutsch}},\ }\bibfield  {title} {\bibinfo {title} {Quantum statistical
  mechanics in a closed system},\ }\href
  {https://doi.org/10.1103/PhysRevA.43.2046} {\bibfield  {journal} {\bibinfo
  {journal} {Phys. Rev. A}\ }\textbf {\bibinfo {volume} {43}},\ \bibinfo
  {pages} {2046} (\bibinfo {year} {1991})}\BibitemShut {NoStop}%
\bibitem [{\citenamefont
  {Srednicki}(1994)}]{srednickiChaosQuantumThermalization1994}%
  \BibitemOpen
  \bibfield  {author} {\bibinfo {author} {\bibfnamefont {M.}~\bibnamefont
  {Srednicki}},\ }\bibfield  {title} {\bibinfo {title} {Chaos and quantum
  thermalization},\ }\href {https://doi.org/10.1103/PhysRevE.50.888} {\bibfield
   {journal} {\bibinfo  {journal} {Phys. Rev. E}\ }\textbf {\bibinfo {volume}
  {50}},\ \bibinfo {pages} {888} (\bibinfo {year} {1994})}\BibitemShut
  {NoStop}%
\bibitem [{\citenamefont
  {Srednicki}(1999)}]{srednickiApproachThermalEquilibrium1999}%
  \BibitemOpen
  \bibfield  {author} {\bibinfo {author} {\bibfnamefont {M.}~\bibnamefont
  {Srednicki}},\ }\bibfield  {title} {\bibinfo {title} {The approach to thermal
  equilibrium in quantized chaotic systems},\ }\href
  {https://doi.org/10.1088/0305-4470/32/7/007} {\bibfield  {journal} {\bibinfo
  {journal} {J. Phys. A: Math. Gen.}\ }\textbf {\bibinfo {volume} {32}},\
  \bibinfo {pages} {1163} (\bibinfo {year} {1999})}\BibitemShut {NoStop}%
\bibitem [{\citenamefont {Nandkishore}\ and\ \citenamefont
  {Huse}(2015)}]{nandkishoreManyBodyLocalizationThermalization2015}%
  \BibitemOpen
  \bibfield  {author} {\bibinfo {author} {\bibfnamefont {R.}~\bibnamefont
  {Nandkishore}}\ and\ \bibinfo {author} {\bibfnamefont {D.~A.}\ \bibnamefont
  {Huse}},\ }\bibfield  {title} {\bibinfo {title} {Many-{{Body Localization}}
  and {{Thermalization}} in {{Quantum Statistical Mechanics}}},\ }\href
  {https://doi.org/10.1146/annurev-conmatphys-031214-014726} {\bibfield
  {journal} {\bibinfo  {journal} {Annu. Rev. Condens. Matter Phys.}\ }\textbf
  {\bibinfo {volume} {6}},\ \bibinfo {pages} {15} (\bibinfo {year}
  {2015})}\BibitemShut {NoStop}%
\bibitem [{\citenamefont {D'Alessio}\ \emph {et~al.}(2016)\citenamefont
  {D'Alessio}, \citenamefont {Kafri}, \citenamefont {Polkovnikov},\ and\
  \citenamefont {Rigol}}]{dalessioQuantumChaosEigenstate2016}%
  \BibitemOpen
  \bibfield  {author} {\bibinfo {author} {\bibfnamefont {L.}~\bibnamefont
  {D'Alessio}}, \bibinfo {author} {\bibfnamefont {Y.}~\bibnamefont {Kafri}},
  \bibinfo {author} {\bibfnamefont {A.}~\bibnamefont {Polkovnikov}},\ and\
  \bibinfo {author} {\bibfnamefont {M.}~\bibnamefont {Rigol}},\ }\bibfield
  {title} {\bibinfo {title} {From quantum chaos and eigenstate thermalization
  to statistical mechanics and thermodynamics},\ }\href
  {https://doi.org/10.1080/00018732.2016.1198134} {\bibfield  {journal}
  {\bibinfo  {journal} {Adv. Phys.}\ }\textbf {\bibinfo {volume} {65}},\
  \bibinfo {pages} {239} (\bibinfo {year} {2016})}\BibitemShut {NoStop}%
\bibitem [{\citenamefont {Gogolin}\ and\ \citenamefont
  {Eisert}(2016)}]{gogolinEquilibrationThermalisationEmergence2016}%
  \BibitemOpen
  \bibfield  {author} {\bibinfo {author} {\bibfnamefont {C.}~\bibnamefont
  {Gogolin}}\ and\ \bibinfo {author} {\bibfnamefont {J.}~\bibnamefont
  {Eisert}},\ }\bibfield  {title} {\bibinfo {title} {Equilibration,
  thermalisation, and the emergence of statistical mechanics in closed quantum
  systems},\ }\href {https://doi.org/10.1088/0034-4885/79/5/056001} {\bibfield
  {journal} {\bibinfo  {journal} {Rep. Prog. Phys.}\ }\textbf {\bibinfo
  {volume} {79}},\ \bibinfo {pages} {056001} (\bibinfo {year}
  {2016})}\BibitemShut {NoStop}%
\bibitem [{\citenamefont
  {Anderson}(1958)}]{andersonAbsenceDiffusionCertain1958}%
  \BibitemOpen
  \bibfield  {author} {\bibinfo {author} {\bibfnamefont {P.~W.}\ \bibnamefont
  {Anderson}},\ }\bibfield  {title} {\bibinfo {title} {Absence of {{Diffusion}}
  in {{Certain Random Lattices}}},\ }\href
  {https://doi.org/10.1103/PhysRev.109.1492} {\bibfield  {journal} {\bibinfo
  {journal} {Phys. Rev.}\ }\textbf {\bibinfo {volume} {109}},\ \bibinfo {pages}
  {1492} (\bibinfo {year} {1958})}\BibitemShut {NoStop}%
\bibitem [{\citenamefont {Mott}\ \emph {et~al.}(1975)\citenamefont {Mott},
  \citenamefont {Pepper}, \citenamefont {Pollitt}, \citenamefont {Wallis},\
  and\ \citenamefont {Adkins}}]{mottAndersonTransition1975}%
  \BibitemOpen
  \bibfield  {author} {\bibinfo {author} {\bibfnamefont {N.~F.}\ \bibnamefont
  {Mott}}, \bibinfo {author} {\bibfnamefont {M.}~\bibnamefont {Pepper}},
  \bibinfo {author} {\bibfnamefont {S.}~\bibnamefont {Pollitt}}, \bibinfo
  {author} {\bibfnamefont {R.~H.}\ \bibnamefont {Wallis}},\ and\ \bibinfo
  {author} {\bibfnamefont {C.~J.}\ \bibnamefont {Adkins}},\ }\bibfield  {title}
  {\bibinfo {title} {The {{Anderson}} transition},\ }\href
  {https://doi.org/10.1098/rspa.1975.0131} {\bibfield  {journal} {\bibinfo
  {journal} {Proc. R. Soc. Lond. A}\ }\textbf {\bibinfo {volume} {345}},\
  \bibinfo {pages} {169} (\bibinfo {year} {1975})}\BibitemShut {NoStop}%
\bibitem [{\citenamefont
  {Adkins}(1978)}]{adkinsThresholdConductionInversion1978}%
  \BibitemOpen
  \bibfield  {author} {\bibinfo {author} {\bibfnamefont {C.~J.}\ \bibnamefont
  {Adkins}},\ }\bibfield  {title} {\bibinfo {title} {Threshold conduction in
  inversion layers},\ }\href {https://doi.org/10.1088/0022-3719/11/5/008}
  {\bibfield  {journal} {\bibinfo  {journal} {J. Phys. C: Solid State Phys.}\
  }\textbf {\bibinfo {volume} {11}},\ \bibinfo {pages} {851} (\bibinfo {year}
  {1978})}\BibitemShut {NoStop}%
\bibitem [{\citenamefont {Abrahams}\ \emph {et~al.}(1979)\citenamefont
  {Abrahams}, \citenamefont {Anderson}, \citenamefont {Licciardello},\ and\
  \citenamefont {Ramakrishnan}}]{abrahamsScalingTheoryLocalization1979}%
  \BibitemOpen
  \bibfield  {author} {\bibinfo {author} {\bibfnamefont {E.}~\bibnamefont
  {Abrahams}}, \bibinfo {author} {\bibfnamefont {P.~W.}\ \bibnamefont
  {Anderson}}, \bibinfo {author} {\bibfnamefont {D.~C.}\ \bibnamefont
  {Licciardello}},\ and\ \bibinfo {author} {\bibfnamefont {T.~V.}\ \bibnamefont
  {Ramakrishnan}},\ }\bibfield  {title} {\bibinfo {title} {Scaling {{Theory}}
  of {{Localization}}: {{Absence}} of {{Quantum Diffusion}} in {{Two
  Dimensions}}},\ }\href {https://doi.org/10.1103/PhysRevLett.42.673}
  {\bibfield  {journal} {\bibinfo  {journal} {Phys. Rev. Lett.}\ }\textbf
  {\bibinfo {volume} {42}},\ \bibinfo {pages} {673} (\bibinfo {year}
  {1979})}\BibitemShut {NoStop}%
\bibitem [{\citenamefont {Evers}\ and\ \citenamefont
  {Mirlin}(2008)}]{eversAndersonTransitions2008}%
  \BibitemOpen
  \bibfield  {author} {\bibinfo {author} {\bibfnamefont {F.}~\bibnamefont
  {Evers}}\ and\ \bibinfo {author} {\bibfnamefont {A.~D.}\ \bibnamefont
  {Mirlin}},\ }\bibfield  {title} {\bibinfo {title} {Anderson transitions},\
  }\href {https://doi.org/10.1103/RevModPhys.80.1355} {\bibfield  {journal}
  {\bibinfo  {journal} {Rev. Mod. Phys.}\ }\textbf {\bibinfo {volume} {80}},\
  \bibinfo {pages} {1355} (\bibinfo {year} {2008})}\BibitemShut {NoStop}%
\bibitem [{\citenamefont {Basko}\ \emph {et~al.}(2006)\citenamefont {Basko},
  \citenamefont {Aleiner},\ and\ \citenamefont
  {Altshuler}}]{baskoMetalinsulatorTransitionWeakly2006}%
  \BibitemOpen
  \bibfield  {author} {\bibinfo {author} {\bibfnamefont {D.~M.}\ \bibnamefont
  {Basko}}, \bibinfo {author} {\bibfnamefont {I.~L.}\ \bibnamefont {Aleiner}},\
  and\ \bibinfo {author} {\bibfnamefont {B.~L.}\ \bibnamefont {Altshuler}},\
  }\bibfield  {title} {\bibinfo {title} {Metal-insulator transition in a weakly
  interacting many-electron system with localized single-particle states},\
  }\href {https://doi.org/10.1016/j.aop.2005.11.014} {\bibfield  {journal}
  {\bibinfo  {journal} {Ann. Phys.}\ }\textbf {\bibinfo {volume} {321}},\
  \bibinfo {pages} {1126} (\bibinfo {year} {2006})}\BibitemShut {NoStop}%
\bibitem [{\citenamefont {Sierant}\ \emph {et~al.}(2025)\citenamefont
  {Sierant}, \citenamefont {Lewenstein}, \citenamefont {Scardicchio},
  \citenamefont {Vidmar},\ and\ \citenamefont
  {Zakrzewski}}]{sierantManybodyLocalizationAge2025}%
  \BibitemOpen
  \bibfield  {author} {\bibinfo {author} {\bibfnamefont {P.}~\bibnamefont
  {Sierant}}, \bibinfo {author} {\bibfnamefont {M.}~\bibnamefont {Lewenstein}},
  \bibinfo {author} {\bibfnamefont {A.}~\bibnamefont {Scardicchio}}, \bibinfo
  {author} {\bibfnamefont {L.}~\bibnamefont {Vidmar}},\ and\ \bibinfo {author}
  {\bibfnamefont {J.}~\bibnamefont {Zakrzewski}},\ }\bibfield  {title}
  {\bibinfo {title} {Many-body localization in the age of classical
  computing*},\ }\href {https://doi.org/10.1088/1361-6633/ad9756} {\bibfield
  {journal} {\bibinfo  {journal} {Rep. Prog. Phys.}\ }\textbf {\bibinfo
  {volume} {88}},\ \bibinfo {pages} {026502} (\bibinfo {year}
  {2025})}\BibitemShut {NoStop}%
\bibitem [{\citenamefont
  {Imbrie}(2016)}]{imbrieManyBodyLocalizationQuantum2016}%
  \BibitemOpen
  \bibfield  {author} {\bibinfo {author} {\bibfnamefont {J.~Z.}\ \bibnamefont
  {Imbrie}},\ }\bibfield  {title} {\bibinfo {title} {On {{Many-Body
  Localization}} for {{Quantum Spin Chains}}},\ }\href
  {https://doi.org/10.1007/s10955-016-1508-x} {\bibfield  {journal} {\bibinfo
  {journal} {J. Stat. Phys.}\ }\textbf {\bibinfo {volume} {163}},\ \bibinfo
  {pages} {998} (\bibinfo {year} {2016})}\BibitemShut {NoStop}%
\bibitem [{\citenamefont {Oganesyan}\ and\ \citenamefont
  {Huse}(2007)}]{oganesyanLocalizationInteractingFermions2007}%
  \BibitemOpen
  \bibfield  {author} {\bibinfo {author} {\bibfnamefont {V.}~\bibnamefont
  {Oganesyan}}\ and\ \bibinfo {author} {\bibfnamefont {D.~A.}\ \bibnamefont
  {Huse}},\ }\bibfield  {title} {\bibinfo {title} {Localization of interacting
  fermions at high temperature},\ }\href
  {https://doi.org/10.1103/PhysRevB.75.155111} {\bibfield  {journal} {\bibinfo
  {journal} {Phys. Rev. B}\ }\textbf {\bibinfo {volume} {75}},\ \bibinfo
  {pages} {155111} (\bibinfo {year} {2007})}\BibitemShut {NoStop}%
\bibitem [{\citenamefont {Pal}\ and\ \citenamefont
  {Huse}(2010)}]{palManybodyLocalizationPhase2010}%
  \BibitemOpen
  \bibfield  {author} {\bibinfo {author} {\bibfnamefont {A.}~\bibnamefont
  {Pal}}\ and\ \bibinfo {author} {\bibfnamefont {D.~A.}\ \bibnamefont {Huse}},\
  }\bibfield  {title} {\bibinfo {title} {Many-body localization phase
  transition},\ }\href {https://doi.org/10.1103/PhysRevB.82.174411} {\bibfield
  {journal} {\bibinfo  {journal} {Phys. Rev. B}\ }\textbf {\bibinfo {volume}
  {82}},\ \bibinfo {pages} {174411} (\bibinfo {year} {2010})}\BibitemShut
  {NoStop}%
\bibitem [{\citenamefont {{\v Z}nidari{\v c}}\ \emph
  {et~al.}(2008)\citenamefont {{\v Z}nidari{\v c}}, \citenamefont {Prosen},\
  and\ \citenamefont {Prelov{\v
  s}ek}}]{znidaricManybodyLocalizationHeisenberg2008}%
  \BibitemOpen
  \bibfield  {author} {\bibinfo {author} {\bibfnamefont {M.}~\bibnamefont {{\v
  Z}nidari{\v c}}}, \bibinfo {author} {\bibfnamefont {T.}~\bibnamefont
  {Prosen}},\ and\ \bibinfo {author} {\bibfnamefont {P.}~\bibnamefont
  {Prelov{\v s}ek}},\ }\bibfield  {title} {\bibinfo {title} {Many-body
  localization in the {{Heisenberg XXZ}} magnet in a random field},\ }\href
  {https://doi.org/10.1103/PhysRevB.77.064426} {\bibfield  {journal} {\bibinfo
  {journal} {Phys. Rev. B}\ }\textbf {\bibinfo {volume} {77}},\ \bibinfo
  {pages} {064426} (\bibinfo {year} {2008})}\BibitemShut {NoStop}%
\bibitem [{\citenamefont {Oganesyan}\ \emph {et~al.}(2009)\citenamefont
  {Oganesyan}, \citenamefont {Pal},\ and\ \citenamefont
  {Huse}}]{oganesyanEnergyTransportDisordered2009}%
  \BibitemOpen
  \bibfield  {author} {\bibinfo {author} {\bibfnamefont {V.}~\bibnamefont
  {Oganesyan}}, \bibinfo {author} {\bibfnamefont {A.}~\bibnamefont {Pal}},\
  and\ \bibinfo {author} {\bibfnamefont {D.~A.}\ \bibnamefont {Huse}},\
  }\bibfield  {title} {\bibinfo {title} {Energy transport in disordered
  classical spin chains},\ }\href {https://doi.org/10.1103/PhysRevB.80.115104}
  {\bibfield  {journal} {\bibinfo  {journal} {Phys. Rev. B}\ }\textbf {\bibinfo
  {volume} {80}},\ \bibinfo {pages} {115104} (\bibinfo {year}
  {2009})}\BibitemShut {NoStop}%
\bibitem [{\citenamefont {Berkelbach}\ and\ \citenamefont
  {Reichman}(2010)}]{berkelbachConductivityDisorderedQuantum2010}%
  \BibitemOpen
  \bibfield  {author} {\bibinfo {author} {\bibfnamefont {T.~C.}\ \bibnamefont
  {Berkelbach}}\ and\ \bibinfo {author} {\bibfnamefont {D.~R.}\ \bibnamefont
  {Reichman}},\ }\bibfield  {title} {\bibinfo {title} {Conductivity of
  disordered quantum lattice models at infinite temperature: {{Many-body}}
  localization},\ }\href {https://doi.org/10.1103/PhysRevB.81.224429}
  {\bibfield  {journal} {\bibinfo  {journal} {Phys. Rev. B}\ }\textbf {\bibinfo
  {volume} {81}},\ \bibinfo {pages} {224429} (\bibinfo {year}
  {2010})}\BibitemShut {NoStop}%
\bibitem [{\citenamefont {O'Brien}\ \emph {et~al.}(2016)\citenamefont
  {O'Brien}, \citenamefont {Abanin}, \citenamefont {Vidal},\ and\ \citenamefont
  {Papi{\'c}}}]{obrienExplicitConstructionLocal2016}%
  \BibitemOpen
  \bibfield  {author} {\bibinfo {author} {\bibfnamefont {T.~E.}\ \bibnamefont
  {O'Brien}}, \bibinfo {author} {\bibfnamefont {D.~A.}\ \bibnamefont {Abanin}},
  \bibinfo {author} {\bibfnamefont {G.}~\bibnamefont {Vidal}},\ and\ \bibinfo
  {author} {\bibfnamefont {Z.}~\bibnamefont {Papi{\'c}}},\ }\bibfield  {title}
  {\bibinfo {title} {Explicit construction of local conserved operators in
  disordered many-body systems},\ }\href
  {https://doi.org/10.1103/PhysRevB.94.144208} {\bibfield  {journal} {\bibinfo
  {journal} {Phys. Rev. B}\ }\textbf {\bibinfo {volume} {94}},\ \bibinfo
  {pages} {144208} (\bibinfo {year} {2016})}\BibitemShut {NoStop}%
\bibitem [{\citenamefont {Serbyn}\ \emph {et~al.}(2015)\citenamefont {Serbyn},
  \citenamefont {Papi{\'c}},\ and\ \citenamefont
  {Abanin}}]{serbynCriterionManyBodyLocalizationDelocalization2015}%
  \BibitemOpen
  \bibfield  {author} {\bibinfo {author} {\bibfnamefont {M.}~\bibnamefont
  {Serbyn}}, \bibinfo {author} {\bibfnamefont {Z.}~\bibnamefont {Papi{\'c}}},\
  and\ \bibinfo {author} {\bibfnamefont {D.~A.}\ \bibnamefont {Abanin}},\
  }\bibfield  {title} {\bibinfo {title} {Criterion for {{Many-Body
  Localization-Delocalization Phase Transition}}},\ }\href
  {https://doi.org/10.1103/PhysRevX.5.041047} {\bibfield  {journal} {\bibinfo
  {journal} {Phys. Rev. X}\ }\textbf {\bibinfo {volume} {5}},\ \bibinfo {pages}
  {041047} (\bibinfo {year} {2015})}\BibitemShut {NoStop}%
\bibitem [{\citenamefont {Bardarson}\ \emph {et~al.}(2012)\citenamefont
  {Bardarson}, \citenamefont {Pollmann},\ and\ \citenamefont
  {Moore}}]{bardarsonUnboundedGrowthEntanglement2012}%
  \BibitemOpen
  \bibfield  {author} {\bibinfo {author} {\bibfnamefont {J.~H.}\ \bibnamefont
  {Bardarson}}, \bibinfo {author} {\bibfnamefont {F.}~\bibnamefont
  {Pollmann}},\ and\ \bibinfo {author} {\bibfnamefont {J.~E.}\ \bibnamefont
  {Moore}},\ }\bibfield  {title} {\bibinfo {title} {Unbounded {{Growth}} of
  {{Entanglement}} in {{Models}} of {{Many-Body Localization}}},\ }\href
  {https://doi.org/10.1103/PhysRevLett.109.017202} {\bibfield  {journal}
  {\bibinfo  {journal} {Phys. Rev. Lett.}\ }\textbf {\bibinfo {volume} {109}},\
  \bibinfo {pages} {017202} (\bibinfo {year} {2012})}\BibitemShut {NoStop}%
\bibitem [{\citenamefont {Serbyn}\ \emph {et~al.}(2013)\citenamefont {Serbyn},
  \citenamefont {Papi{\'c}},\ and\ \citenamefont
  {Abanin}}]{serbynLocalConservationLaws2013}%
  \BibitemOpen
  \bibfield  {author} {\bibinfo {author} {\bibfnamefont {M.}~\bibnamefont
  {Serbyn}}, \bibinfo {author} {\bibfnamefont {Z.}~\bibnamefont {Papi{\'c}}},\
  and\ \bibinfo {author} {\bibfnamefont {D.~A.}\ \bibnamefont {Abanin}},\
  }\bibfield  {title} {\bibinfo {title} {Local {{Conservation Laws}} and the
  {{Structure}} of the {{Many-Body Localized States}}},\ }\href
  {https://doi.org/10.1103/PhysRevLett.111.127201} {\bibfield  {journal}
  {\bibinfo  {journal} {Phys. Rev. Lett.}\ }\textbf {\bibinfo {volume} {111}},\
  \bibinfo {pages} {127201} (\bibinfo {year} {2013})}\BibitemShut {NoStop}%
\bibitem [{\citenamefont {Vasseur}\ \emph {et~al.}(2015)\citenamefont
  {Vasseur}, \citenamefont {Potter},\ and\ \citenamefont
  {Parameswaran}}]{vasseurQuantumCriticalityHot2015}%
  \BibitemOpen
  \bibfield  {author} {\bibinfo {author} {\bibfnamefont {R.}~\bibnamefont
  {Vasseur}}, \bibinfo {author} {\bibfnamefont {A.~C.}\ \bibnamefont
  {Potter}},\ and\ \bibinfo {author} {\bibfnamefont {S.~A.}\ \bibnamefont
  {Parameswaran}},\ }\bibfield  {title} {\bibinfo {title} {Quantum
  {{Criticality}} of {{Hot Random Spin Chains}}},\ }\href
  {https://doi.org/10.1103/PhysRevLett.114.217201} {\bibfield  {journal}
  {\bibinfo  {journal} {Phys. Rev. Lett.}\ }\textbf {\bibinfo {volume} {114}},\
  \bibinfo {pages} {217201} (\bibinfo {year} {2015})}\BibitemShut {NoStop}%
\bibitem [{\citenamefont {Doggen}\ \emph {et~al.}(2021)\citenamefont {Doggen},
  \citenamefont {Gornyi}, \citenamefont {Mirlin},\ and\ \citenamefont
  {Polyakov}}]{doggenManybodyLocalizationLarge2021}%
  \BibitemOpen
  \bibfield  {author} {\bibinfo {author} {\bibfnamefont {E.~V.~H.}\
  \bibnamefont {Doggen}}, \bibinfo {author} {\bibfnamefont {I.~V.}\
  \bibnamefont {Gornyi}}, \bibinfo {author} {\bibfnamefont {A.~D.}\
  \bibnamefont {Mirlin}},\ and\ \bibinfo {author} {\bibfnamefont {D.~G.}\
  \bibnamefont {Polyakov}},\ }\bibfield  {title} {\bibinfo {title} {Many-body
  localization in large systems: {{Matrix-product-state}} approach},\ }\href
  {https://doi.org/10.1016/j.aop.2021.168437} {\bibfield  {journal} {\bibinfo
  {journal} {Ann. Phys.}\ }\bibinfo {series} {Special {{Issue}} on
  {{Localisation}} 2020},\ \textbf {\bibinfo {volume} {435}},\ \bibinfo {pages}
  {168437} (\bibinfo {year} {2021})}\BibitemShut {NoStop}%
\bibitem [{\citenamefont {Devakul}\ and\ \citenamefont
  {Singh}(2015)}]{devakulEarlyBreakdownAreaLaw2015}%
  \BibitemOpen
  \bibfield  {author} {\bibinfo {author} {\bibfnamefont {T.}~\bibnamefont
  {Devakul}}\ and\ \bibinfo {author} {\bibfnamefont {R.~R.~P.}\ \bibnamefont
  {Singh}},\ }\bibfield  {title} {\bibinfo {title} {Early {{Breakdown}} of
  {{Area-Law Entanglement}} at the {{Many-Body Delocalization Transition}}},\
  }\href {https://doi.org/10.1103/PhysRevLett.115.187201} {\bibfield  {journal}
  {\bibinfo  {journal} {Phys. Rev. Lett.}\ }\textbf {\bibinfo {volume} {115}},\
  \bibinfo {pages} {187201} (\bibinfo {year} {2015})}\BibitemShut {NoStop}%
\bibitem [{\citenamefont {Mac{\'e}}\ \emph {et~al.}(2019)\citenamefont
  {Mac{\'e}}, \citenamefont {Alet},\ and\ \citenamefont
  {Laflorencie}}]{maceMultifractalScalingsManyBody2019}%
  \BibitemOpen
  \bibfield  {author} {\bibinfo {author} {\bibfnamefont {N.}~\bibnamefont
  {Mac{\'e}}}, \bibinfo {author} {\bibfnamefont {F.}~\bibnamefont {Alet}},\
  and\ \bibinfo {author} {\bibfnamefont {N.}~\bibnamefont {Laflorencie}},\
  }\bibfield  {title} {\bibinfo {title} {Multifractal {{Scalings Across}} the
  {{Many-Body Localization Transition}}},\ }\href
  {https://doi.org/10.1103/PhysRevLett.123.180601} {\bibfield  {journal}
  {\bibinfo  {journal} {Phys. Rev. Lett.}\ }\textbf {\bibinfo {volume} {123}},\
  \bibinfo {pages} {180601} (\bibinfo {year} {2019})}\BibitemShut {NoStop}%
\bibitem [{\citenamefont {Sierant}\ \emph {et~al.}(2020)\citenamefont
  {Sierant}, \citenamefont {Lewenstein},\ and\ \citenamefont
  {Zakrzewski}}]{sierantPolynomiallyFilteredExact2020}%
  \BibitemOpen
  \bibfield  {author} {\bibinfo {author} {\bibfnamefont {P.}~\bibnamefont
  {Sierant}}, \bibinfo {author} {\bibfnamefont {M.}~\bibnamefont
  {Lewenstein}},\ and\ \bibinfo {author} {\bibfnamefont {J.}~\bibnamefont
  {Zakrzewski}},\ }\bibfield  {title} {\bibinfo {title} {Polynomially
  {{Filtered Exact Diagonalization Approach}} to {{Many-Body Localization}}},\
  }\href {https://doi.org/10.1103/PhysRevLett.125.156601} {\bibfield  {journal}
  {\bibinfo  {journal} {Phys. Rev. Lett.}\ }\textbf {\bibinfo {volume} {125}},\
  \bibinfo {pages} {156601} (\bibinfo {year} {2020})}\BibitemShut {NoStop}%
\bibitem [{\citenamefont {Luitz}\ \emph {et~al.}(2015)\citenamefont {Luitz},
  \citenamefont {Laflorencie},\ and\ \citenamefont
  {Alet}}]{luitzManybodyLocalizationEdge2015}%
  \BibitemOpen
  \bibfield  {author} {\bibinfo {author} {\bibfnamefont {D.~J.}\ \bibnamefont
  {Luitz}}, \bibinfo {author} {\bibfnamefont {N.}~\bibnamefont {Laflorencie}},\
  and\ \bibinfo {author} {\bibfnamefont {F.}~\bibnamefont {Alet}},\ }\bibfield
  {title} {\bibinfo {title} {Many-body localization edge in the random-field
  {{Heisenberg}} chain},\ }\href {https://doi.org/10.1103/PhysRevB.91.081103}
  {\bibfield  {journal} {\bibinfo  {journal} {Phys. Rev. B}\ }\textbf {\bibinfo
  {volume} {91}},\ \bibinfo {pages} {081103} (\bibinfo {year}
  {2015})}\BibitemShut {NoStop}%
\bibitem [{\citenamefont {Doggen}\ \emph {et~al.}(2018)\citenamefont {Doggen},
  \citenamefont {Schindler}, \citenamefont {Tikhonov}, \citenamefont {Mirlin},
  \citenamefont {Neupert}, \citenamefont {Polyakov},\ and\ \citenamefont
  {Gornyi}}]{doggenManybodyLocalizationDelocalization2018}%
  \BibitemOpen
  \bibfield  {author} {\bibinfo {author} {\bibfnamefont {E.~V.~H.}\
  \bibnamefont {Doggen}}, \bibinfo {author} {\bibfnamefont {F.}~\bibnamefont
  {Schindler}}, \bibinfo {author} {\bibfnamefont {K.~S.}\ \bibnamefont
  {Tikhonov}}, \bibinfo {author} {\bibfnamefont {A.~D.}\ \bibnamefont
  {Mirlin}}, \bibinfo {author} {\bibfnamefont {T.}~\bibnamefont {Neupert}},
  \bibinfo {author} {\bibfnamefont {D.~G.}\ \bibnamefont {Polyakov}},\ and\
  \bibinfo {author} {\bibfnamefont {I.~V.}\ \bibnamefont {Gornyi}},\ }\bibfield
   {title} {\bibinfo {title} {Many-body localization and delocalization in
  large quantum chains},\ }\href {https://doi.org/10.1103/PhysRevB.98.174202}
  {\bibfield  {journal} {\bibinfo  {journal} {Phys. Rev. B}\ }\textbf {\bibinfo
  {volume} {98}},\ \bibinfo {pages} {174202} (\bibinfo {year}
  {2018})}\BibitemShut {NoStop}%
\bibitem [{\citenamefont {Evers}\ \emph {et~al.}(2023)\citenamefont {Evers},
  \citenamefont {Modak},\ and\ \citenamefont
  {Bera}}]{eversInternalClockManybody2023}%
  \BibitemOpen
  \bibfield  {author} {\bibinfo {author} {\bibfnamefont {F.}~\bibnamefont
  {Evers}}, \bibinfo {author} {\bibfnamefont {I.}~\bibnamefont {Modak}},\ and\
  \bibinfo {author} {\bibfnamefont {S.}~\bibnamefont {Bera}},\ }\bibfield
  {title} {\bibinfo {title} {Internal clock of many-body delocalization},\
  }\href {https://doi.org/10.1103/PhysRevB.108.134204} {\bibfield  {journal}
  {\bibinfo  {journal} {Phys. Rev. B}\ }\textbf {\bibinfo {volume} {108}},\
  \bibinfo {pages} {134204} (\bibinfo {year} {2023})}\BibitemShut {NoStop}%
\bibitem [{\citenamefont {Sels}\ and\ \citenamefont
  {Polkovnikov}(2023)}]{selsThermalizationDiluteImpurities2023}%
  \BibitemOpen
  \bibfield  {author} {\bibinfo {author} {\bibfnamefont {D.}~\bibnamefont
  {Sels}}\ and\ \bibinfo {author} {\bibfnamefont {A.}~\bibnamefont
  {Polkovnikov}},\ }\bibfield  {title} {\bibinfo {title} {Thermalization of
  {{Dilute Impurities}} in {{One-Dimensional Spin Chains}}},\ }\href
  {https://doi.org/10.1103/PhysRevX.13.011041} {\bibfield  {journal} {\bibinfo
  {journal} {Phys. Rev. X}\ }\textbf {\bibinfo {volume} {13}},\ \bibinfo
  {pages} {011041} (\bibinfo {year} {2023})}\BibitemShut {NoStop}%
\bibitem [{\citenamefont {Colmenarez}\ \emph {et~al.}(2019)\citenamefont
  {Colmenarez}, \citenamefont {McClarty}, \citenamefont {Haque},\ and\
  \citenamefont {Luitz}}]{colmenarezStatisticsCorrelationFunctions2019}%
  \BibitemOpen
  \bibfield  {author} {\bibinfo {author} {\bibfnamefont {L.~A.}\ \bibnamefont
  {Colmenarez}}, \bibinfo {author} {\bibfnamefont {P.~A.}\ \bibnamefont
  {McClarty}}, \bibinfo {author} {\bibfnamefont {M.}~\bibnamefont {Haque}},\
  and\ \bibinfo {author} {\bibfnamefont {D.~J.}\ \bibnamefont {Luitz}},\
  }\bibfield  {title} {\bibinfo {title} {Statistics of correlation functions in
  the random {{Heisenberg}} chain},\ }\href
  {https://doi.org/10.21468/SciPostPhys.7.5.064} {\bibfield  {journal}
  {\bibinfo  {journal} {SciPost Phys.}\ }\textbf {\bibinfo {volume} {7}},\
  \bibinfo {pages} {064} (\bibinfo {year} {2019})}\BibitemShut {NoStop}%
\bibitem [{\citenamefont {Pietracaprina}\ \emph {et~al.}(2018)\citenamefont
  {Pietracaprina}, \citenamefont {Mac{\'e}}, \citenamefont {Luitz},\ and\
  \citenamefont {Alet}}]{pietracaprinaShiftinvertDiagonalizationLarge2018}%
  \BibitemOpen
  \bibfield  {author} {\bibinfo {author} {\bibfnamefont {F.}~\bibnamefont
  {Pietracaprina}}, \bibinfo {author} {\bibfnamefont {N.}~\bibnamefont
  {Mac{\'e}}}, \bibinfo {author} {\bibfnamefont {D.~J.}\ \bibnamefont
  {Luitz}},\ and\ \bibinfo {author} {\bibfnamefont {F.}~\bibnamefont {Alet}},\
  }\bibfield  {title} {\bibinfo {title} {Shift-invert diagonalization of large
  many-body localizing spin chains},\ }\href
  {https://doi.org/10.21468/SciPostPhys.5.5.045} {\bibfield  {journal}
  {\bibinfo  {journal} {SciPost Phys.}\ }\textbf {\bibinfo {volume} {5}},\
  \bibinfo {pages} {045} (\bibinfo {year} {2018})}\BibitemShut {NoStop}%
\bibitem [{\citenamefont {Protopopov}\ \emph {et~al.}(2017)\citenamefont
  {Protopopov}, \citenamefont {Ho},\ and\ \citenamefont
  {Abanin}}]{protopopovEffectSU2Symmetry2017}%
  \BibitemOpen
  \bibfield  {author} {\bibinfo {author} {\bibfnamefont {I.~V.}\ \bibnamefont
  {Protopopov}}, \bibinfo {author} {\bibfnamefont {W.~W.}\ \bibnamefont {Ho}},\
  and\ \bibinfo {author} {\bibfnamefont {D.~A.}\ \bibnamefont {Abanin}},\
  }\bibfield  {title} {\bibinfo {title} {Effect of {{SU}}(2) symmetry on
  many-body localization and thermalization},\ }\href
  {https://doi.org/10.1103/PhysRevB.96.041122} {\bibfield  {journal} {\bibinfo
  {journal} {Phys. Rev. B}\ }\textbf {\bibinfo {volume} {96}},\ \bibinfo
  {pages} {041122} (\bibinfo {year} {2017})}\BibitemShut {NoStop}%
\bibitem [{\citenamefont {Prelov{\v s}ek}\ \emph {et~al.}(2016)\citenamefont
  {Prelov{\v s}ek}, \citenamefont {Bari{\v s}i{\'c}},\ and\ \citenamefont {{\v
  Z}nidari{\v c}}}]{prelovsekAbsenceFullManybody2016}%
  \BibitemOpen
  \bibfield  {author} {\bibinfo {author} {\bibfnamefont {P.}~\bibnamefont
  {Prelov{\v s}ek}}, \bibinfo {author} {\bibfnamefont {O.~S.}\ \bibnamefont
  {Bari{\v s}i{\'c}}},\ and\ \bibinfo {author} {\bibfnamefont {M.}~\bibnamefont
  {{\v Z}nidari{\v c}}},\ }\bibfield  {title} {\bibinfo {title} {Absence of
  full many-body localization in the disordered {{Hubbard}} chain},\ }\href
  {https://doi.org/10.1103/PhysRevB.94.241104} {\bibfield  {journal} {\bibinfo
  {journal} {Phys. Rev. B}\ }\textbf {\bibinfo {volume} {94}},\ \bibinfo
  {pages} {241104} (\bibinfo {year} {2016})}\BibitemShut {NoStop}%
\bibitem [{\citenamefont {Potter}\ and\ \citenamefont
  {Vasseur}(2016)}]{potterSymmetryConstraintsManybody2016}%
  \BibitemOpen
  \bibfield  {author} {\bibinfo {author} {\bibfnamefont {A.~C.}\ \bibnamefont
  {Potter}}\ and\ \bibinfo {author} {\bibfnamefont {R.}~\bibnamefont
  {Vasseur}},\ }\bibfield  {title} {\bibinfo {title} {Symmetry constraints on
  many-body localization},\ }\href {https://doi.org/10.1103/PhysRevB.94.224206}
  {\bibfield  {journal} {\bibinfo  {journal} {Phys. Rev. B}\ }\textbf {\bibinfo
  {volume} {94}},\ \bibinfo {pages} {224206} (\bibinfo {year}
  {2016})}\BibitemShut {NoStop}%
\bibitem [{\citenamefont
  {Thomson}(2023)}]{thomsonDisorderinducedSpinchargeSeparation2023}%
  \BibitemOpen
  \bibfield  {author} {\bibinfo {author} {\bibfnamefont {S.~J.}\ \bibnamefont
  {Thomson}},\ }\bibfield  {title} {\bibinfo {title} {Disorder-induced
  spin-charge separation in the one-dimensional {{Hubbard}} model},\ }\href
  {https://doi.org/10.1103/PhysRevB.107.L180201} {\bibfield  {journal}
  {\bibinfo  {journal} {Phys. Rev. B}\ }\textbf {\bibinfo {volume} {107}},\
  \bibinfo {pages} {L180201} (\bibinfo {year} {2023})}\BibitemShut {NoStop}%
\bibitem [{\citenamefont {Majidy}\ \emph {et~al.}(2023)\citenamefont {Majidy},
  \citenamefont {Lasek}, \citenamefont {Huse},\ and\ \citenamefont
  {Halpern}}]{majidyNonAbelianSymmetryCan2023}%
  \BibitemOpen
  \bibfield  {author} {\bibinfo {author} {\bibfnamefont {S.}~\bibnamefont
  {Majidy}}, \bibinfo {author} {\bibfnamefont {A.}~\bibnamefont {Lasek}},
  \bibinfo {author} {\bibfnamefont {D.~A.}\ \bibnamefont {Huse}},\ and\
  \bibinfo {author} {\bibfnamefont {N.~Y.}\ \bibnamefont {Halpern}},\
  }\bibfield  {title} {\bibinfo {title} {Non-{{Abelian}} symmetry can increase
  entanglement entropy},\ }\href {https://doi.org/10.1103/PhysRevB.107.045102}
  {\bibfield  {journal} {\bibinfo  {journal} {Phys. Rev. B}\ }\textbf {\bibinfo
  {volume} {107}},\ \bibinfo {pages} {045102} (\bibinfo {year} {2023})},\
  \Eprint {https://arxiv.org/abs/2209.14303} {2209.14303 [cond-mat,
  physics:hep-th, physics:quant-ph]} \BibitemShut {NoStop}%
\bibitem [{\citenamefont {Protopopov}\ \emph {et~al.}(2020)\citenamefont
  {Protopopov}, \citenamefont {Panda}, \citenamefont {Parolini}, \citenamefont
  {Scardicchio}, \citenamefont {Demler},\ and\ \citenamefont
  {Abanin}}]{protopopovNonAbelianSymmetriesDisorder2020}%
  \BibitemOpen
  \bibfield  {author} {\bibinfo {author} {\bibfnamefont {I.~V.}\ \bibnamefont
  {Protopopov}}, \bibinfo {author} {\bibfnamefont {R.~K.}\ \bibnamefont
  {Panda}}, \bibinfo {author} {\bibfnamefont {T.}~\bibnamefont {Parolini}},
  \bibinfo {author} {\bibfnamefont {A.}~\bibnamefont {Scardicchio}}, \bibinfo
  {author} {\bibfnamefont {E.}~\bibnamefont {Demler}},\ and\ \bibinfo {author}
  {\bibfnamefont {D.~A.}\ \bibnamefont {Abanin}},\ }\bibfield  {title}
  {\bibinfo {title} {Non-{{Abelian Symmetries}} and {{Disorder}}: {{A Broad
  Nonergodic Regime}} and {{Anomalous Thermalization}}},\ }\href
  {https://doi.org/10.1103/PhysRevX.10.011025} {\bibfield  {journal} {\bibinfo
  {journal} {Phys. Rev. X}\ }\textbf {\bibinfo {volume} {10}},\ \bibinfo
  {pages} {011025} (\bibinfo {year} {2020})}\BibitemShut {NoStop}%
\bibitem [{\citenamefont {Kozarzewski}\ \emph {et~al.}(2018)\citenamefont
  {Kozarzewski}, \citenamefont {Prelov{\v s}ek},\ and\ \citenamefont
  {Mierzejewski}}]{kozarzewskiSpinSubdiffusionDisordered2018}%
  \BibitemOpen
  \bibfield  {author} {\bibinfo {author} {\bibfnamefont {M.}~\bibnamefont
  {Kozarzewski}}, \bibinfo {author} {\bibfnamefont {P.}~\bibnamefont {Prelov{\v
  s}ek}},\ and\ \bibinfo {author} {\bibfnamefont {M.}~\bibnamefont
  {Mierzejewski}},\ }\bibfield  {title} {\bibinfo {title} {Spin
  {{Subdiffusion}} in the {{Disordered Hubbard Chain}}},\ }\href
  {https://doi.org/10.1103/PhysRevLett.120.246602} {\bibfield  {journal}
  {\bibinfo  {journal} {Phys. Rev. Lett.}\ }\textbf {\bibinfo {volume} {120}},\
  \bibinfo {pages} {246602} (\bibinfo {year} {2018})}\BibitemShut {NoStop}%
\bibitem [{\citenamefont {Murthy}\ \emph {et~al.}(2023)\citenamefont {Murthy},
  \citenamefont {Babakhani}, \citenamefont {Iniguez}, \citenamefont
  {Srednicki},\ and\ \citenamefont
  {Yunger~Halpern}}]{murthyNonAbelianEigenstateThermalization2023}%
  \BibitemOpen
  \bibfield  {author} {\bibinfo {author} {\bibfnamefont {C.}~\bibnamefont
  {Murthy}}, \bibinfo {author} {\bibfnamefont {A.}~\bibnamefont {Babakhani}},
  \bibinfo {author} {\bibfnamefont {F.}~\bibnamefont {Iniguez}}, \bibinfo
  {author} {\bibfnamefont {M.}~\bibnamefont {Srednicki}},\ and\ \bibinfo
  {author} {\bibfnamefont {N.}~\bibnamefont {Yunger~Halpern}},\ }\bibfield
  {title} {\bibinfo {title} {Non-{{Abelian Eigenstate Thermalization
  Hypothesis}}},\ }\href {https://doi.org/10.1103/PhysRevLett.130.140402}
  {\bibfield  {journal} {\bibinfo  {journal} {Phys. Rev. Lett.}\ }\textbf
  {\bibinfo {volume} {130}},\ \bibinfo {pages} {140402} (\bibinfo {year}
  {2023})}\BibitemShut {NoStop}%
\bibitem [{\citenamefont {Siegl}\ and\ \citenamefont
  {Schliemann}(2023)}]{sieglImperfectManyBodyLocalization2023}%
  \BibitemOpen
  \bibfield  {author} {\bibinfo {author} {\bibfnamefont {J.}~\bibnamefont
  {Siegl}}\ and\ \bibinfo {author} {\bibfnamefont {J.}~\bibnamefont
  {Schliemann}},\ }\bibfield  {title} {\bibinfo {title} {Imperfect {{Many-Body
  Localization}} in {{Exchange-Disordered Isotropic Spin Chains}}},\ }\href
  {https://doi.org/10.1088/1367-2630/ad0e1b} {\bibfield  {journal} {\bibinfo
  {journal} {New J. Phys.}\ }\textbf {\bibinfo {volume} {25}},\ \bibinfo
  {pages} {123002} (\bibinfo {year} {2023})}\BibitemShut {NoStop}%
\bibitem [{\citenamefont {Gao}\ and\ \citenamefont
  {R{\"o}mer}(2025)}]{gaoSpectralEntanglementProperties2025}%
  \BibitemOpen
  \bibfield  {author} {\bibinfo {author} {\bibfnamefont {Y.}~\bibnamefont
  {Gao}}\ and\ \bibinfo {author} {\bibfnamefont {R.~A.}\ \bibnamefont
  {R{\"o}mer}},\ }\bibfield  {title} {\bibinfo {title} {Spectral and
  entanglement properties of the random-exchange {{Heisenberg}} chain},\ }\href
  {https://doi.org/10.1103/PhysRevB.111.104202} {\bibfield  {journal} {\bibinfo
   {journal} {Phys. Rev. B}\ }\textbf {\bibinfo {volume} {111}},\ \bibinfo
  {pages} {104202} (\bibinfo {year} {2025})}\BibitemShut {NoStop}%
\bibitem [{\citenamefont {Bera}\ \emph {et~al.}(2015)\citenamefont {Bera},
  \citenamefont {Schomerus}, \citenamefont {{Heidrich-Meisner}},\ and\
  \citenamefont {Bardarson}}]{beraManyBodyLocalizationCharacterized2015}%
  \BibitemOpen
  \bibfield  {author} {\bibinfo {author} {\bibfnamefont {S.}~\bibnamefont
  {Bera}}, \bibinfo {author} {\bibfnamefont {H.}~\bibnamefont {Schomerus}},
  \bibinfo {author} {\bibfnamefont {F.}~\bibnamefont {{Heidrich-Meisner}}},\
  and\ \bibinfo {author} {\bibfnamefont {J.~H.}\ \bibnamefont {Bardarson}},\
  }\bibfield  {title} {\bibinfo {title} {Many-{{Body Localization
  Characterized}} from a {{One-Particle Perspective}}},\ }\href
  {https://doi.org/10.1103/PhysRevLett.115.046603} {\bibfield  {journal}
  {\bibinfo  {journal} {Phys. Rev. Lett.}\ }\textbf {\bibinfo {volume} {115}},\
  \bibinfo {pages} {046603} (\bibinfo {year} {2015})}\BibitemShut {NoStop}%
\bibitem [{\citenamefont {Lin}\ \emph {et~al.}(2018)\citenamefont {Lin},
  \citenamefont {Sbierski}, \citenamefont {Dorfner}, \citenamefont {Karrasch},\
  and\ \citenamefont
  {{Heidrich-Meisner}}}]{linManybodyLocalizationSpinless2018}%
  \BibitemOpen
  \bibfield  {author} {\bibinfo {author} {\bibfnamefont {S.-H.}\ \bibnamefont
  {Lin}}, \bibinfo {author} {\bibfnamefont {B.}~\bibnamefont {Sbierski}},
  \bibinfo {author} {\bibfnamefont {F.}~\bibnamefont {Dorfner}}, \bibinfo
  {author} {\bibfnamefont {C.}~\bibnamefont {Karrasch}},\ and\ \bibinfo
  {author} {\bibfnamefont {F.}~\bibnamefont {{Heidrich-Meisner}}},\ }\bibfield
  {title} {\bibinfo {title} {Many-body localization of spinless fermions with
  attractive interactions in one dimension},\ }\href
  {https://doi.org/10.21468/SciPostPhys.4.1.002} {\bibfield  {journal}
  {\bibinfo  {journal} {SciPost Phys.}\ }\textbf {\bibinfo {volume} {4}},\
  \bibinfo {pages} {002} (\bibinfo {year} {2018})}\BibitemShut {NoStop}%
\bibitem [{\citenamefont {Carrasquilla}\ and\ \citenamefont
  {Melko}(2017)}]{carrasquillaMachineLearningPhases2017}%
  \BibitemOpen
  \bibfield  {author} {\bibinfo {author} {\bibfnamefont {J.}~\bibnamefont
  {Carrasquilla}}\ and\ \bibinfo {author} {\bibfnamefont {R.~G.}\ \bibnamefont
  {Melko}},\ }\bibfield  {title} {\bibinfo {title} {Machine learning phases of
  matter},\ }\href {https://doi.org/10.1038/nphys4035} {\bibfield  {journal}
  {\bibinfo  {journal} {Nature Phys.}\ }\textbf {\bibinfo {volume} {13}},\
  \bibinfo {pages} {431} (\bibinfo {year} {2017})}\BibitemShut {NoStop}%
\bibitem [{\citenamefont {Beach}\ \emph {et~al.}(2018)\citenamefont {Beach},
  \citenamefont {Golubeva},\ and\ \citenamefont
  {Melko}}]{beachMachineLearningVortices2018}%
  \BibitemOpen
  \bibfield  {author} {\bibinfo {author} {\bibfnamefont {M.~J.~S.}\
  \bibnamefont {Beach}}, \bibinfo {author} {\bibfnamefont {A.}~\bibnamefont
  {Golubeva}},\ and\ \bibinfo {author} {\bibfnamefont {R.~G.}\ \bibnamefont
  {Melko}},\ }\bibfield  {title} {\bibinfo {title} {Machine learning vortices
  at the {{Kosterlitz-Thouless}} transition},\ }\href
  {https://doi.org/10.1103/PhysRevB.97.045207} {\bibfield  {journal} {\bibinfo
  {journal} {Phys. Rev. B}\ }\textbf {\bibinfo {volume} {97}},\ \bibinfo
  {pages} {045207} (\bibinfo {year} {2018})}\BibitemShut {NoStop}%
\bibitem [{\citenamefont {Vasseur}\ \emph {et~al.}(2016)\citenamefont
  {Vasseur}, \citenamefont {Friedman}, \citenamefont {Parameswaran},\ and\
  \citenamefont {Potter}}]{vasseurParticleholeSymmetryManybody2016}%
  \BibitemOpen
  \bibfield  {author} {\bibinfo {author} {\bibfnamefont {R.}~\bibnamefont
  {Vasseur}}, \bibinfo {author} {\bibfnamefont {A.~J.}\ \bibnamefont
  {Friedman}}, \bibinfo {author} {\bibfnamefont {S.~A.}\ \bibnamefont
  {Parameswaran}},\ and\ \bibinfo {author} {\bibfnamefont {A.~C.}\ \bibnamefont
  {Potter}},\ }\bibfield  {title} {\bibinfo {title} {Particle-hole symmetry,
  many-body localization, and topological edge modes},\ }\href
  {https://doi.org/10.1103/PhysRevB.93.134207} {\bibfield  {journal} {\bibinfo
  {journal} {Phys. Rev. B}\ }\textbf {\bibinfo {volume} {93}},\ \bibinfo
  {pages} {134207} (\bibinfo {year} {2016})}\BibitemShut {NoStop}%
\end{thebibliography}%
\end{document}